\documentclass[11pt,draftcls,onecolumn,peerreview]{IEEEtran} 
\usepackage{amsmath,amssymb,epsfig,color,graphicx,subfigure,psfrag}

\usepackage{cite}
\newcommand{\NEW}[1]{{{\color{black}#1}}}
\newcommand{\NEWa}[1]{{{\color{black}#1}}}

\newcommand{\seye}{\mathbf{I}}		
\newcommand{\sE}{\mathbb{E}}					
\newcommand{\sent}{\mathbb{H}}
\newcommand{\sO}{\mathcal{O}}		
\newcommand{\su}{\text{u}}	
\newcommand{\shamming}{{d}_\text{H}}
\newcommand{\strans}{^\text{T}}	
\newcommand{\sherm}{^\text{H}}	
\newcommand{\sfrob}{_\text{F}}		
\newcommand{\sexp}{e}							
\newcommand{\sProb}{\text{P}}					

\newcommand{\si}{i}					
\newcommand{\sq}{q}					
\newcommand{\sk}{k}					
\newcommand{\sn}{n}					
\newcommand{\sj}{l}					

\newcommand{\sprob}{f}	
\newcommand{\spmf}{p}	
\newcommand{\sltwo}{\ell^2}
\newcommand{\slinf}{\ell^\infty}

\newcommand{\sri}{y}
\newcommand{\sr}{\mathbf{\sri}}

\newcommand{\syi}{x}
\newcommand{\sy}{\mathbf{\syi}}

\newcommand{\sxi}{x}
\newcommand{\sx}{\mathbf{\sxi}}

\newcommand{\shxi}{\hat{x}}
\newcommand{\shx}{\mathbf{\shxi}}

\newcommand{\swi}{v}
\newcommand{\sw}{\mathbf{\swi}}
\newcommand{\ssigw}{\sigma^2_\swi}

\newcommand{\sHi}{H}
\newcommand{\sH}{\mathbf{\sHi}}
\newcommand{\sh}{\mathbf{h}}

\newcommand{\sMt}{M_\text{T}}		
\newcommand{\sMr}{M_\text{R}}		

\newcommand{\sbit}{b}					
\newcommand{\sbithat}{\hat{\sbit}}					
\newcommand{\scbit}{c}					
\newcommand{\scik}{\scbit_\sk^{(\si)}}	
\newcommand{\scj}{\scbit_{\sj}}	

\newcommand{\shcbit}{\hat{\scbit}}	
\newcommand{\svc}{\mathbf{\scbit}}	
\newcommand{\svhc}{\mathbf{\shcbit}}	
\newcommand{\shcj}{\shcbit_{\sj}}	
\newcommand{\shcik}{\shcbit_\sk^{(\si)}}	

\newcommand{\sN}{N}					

\newcommand{\sA}{\mathcal{A}}
\newcommand{\sQuant}{\mathcal{Q}}		
\newcommand{\smu}{\mu}				
\newcommand{\sM}{|\sA|}					
\newcommand{\sm}{Q}					

\newcommand{\sPt}{E_s}

\newcommand{\sZint}{\mathbb{Z}}

\newcommand{\sLLR}{\Lambda}			
\newcommand{\sLLRik}{\sLLR_{\sk}^{(\si)}}	
\newcommand{\sLLRj}{\sLLR_{\sj}}	
\newcommand{\shLLR}{\tilde{\sLLR}}			
\newcommand{\shLLRj}{{\shLLR_{\sj}}}	
\newcommand{\shLLRik}{\shLLR_{\sk}^{(\si)}}	

\newcommand{\sL}{\mathcal{L}}
\newcommand{\sLcand}{|\sL|}
\newcommand{\sLLRLSD}{\check{\Lambda}}			

\newcommand{\sDham}{D}							
\newcommand{\sdi}{d}							

\newcommand{\shxml}{\shx_\text{ML}}

\newcommand{\sprefac}{\sigma_k^2}
\newcommand{\sv}{\mathbf{\tilde v}}
\newcommand{\sRw}{\mathbf{R}_\sv}			
\newcommand{\syzf}{\hat{\sy}_\text{ZF}}

\newcommand{\symmse}{\hat{\sy}_\text{MMSE}}
\newcommand{\sWi}{W}	
\newcommand{\sWkk}{\sWi_{\sk}}	
\newcommand{\sW}{\mathbf{\sWi}}	

\newcommand{\sPkj}{P_\sk^{(\siter)}}		
\newcommand{\sxsoft}{\tilde{\sxi}}			

\newcommand{\sQqr}{\mathbf{Q}}	
\newcommand{\sRqr}{\mathbf{R}}	
\newcommand{\shxlinf}{\shx_{\infty}}

\newcommand{\sI}{I}			
\newcommand{\sC}{C}			
\newcommand{\sR}{C}			
\newcommand{\sRbar}{R}			
\newcommand{\sReps}{\sC_\epsilon}
\newcommand{\sRH}{\sRbar_\sH}			

\newcommand{\slambda}{R_0}
\newcommand{\shA}{{{\sA}^{\sMt}}}	
\newcommand{\shAcj}{\mathcal{X}^b_{ \sj }}
\newcommand{\shAz}{\mathcal{X}^0_{ \sj }}
\newcommand{\shAo}{\mathcal{X}^1_{ \sj }}	
\newcommand{\sSNR}{\rho}	   %

\newcommand{\sP}{\text{P}}	   %
\newcommand{\spcross}{p_0}	   %

\newcommand{\sPout}{P_\text{out}}
\newcommand{\define}{\triangleq} 

\newcommand{\szi}{z}
\newcommand{\sz}{\mathbf{\szi}}
\newcommand{\sHtilde}{\tilde{\sH}}
\newcommand{\sT}{\mathbf{T}}

\newcommand{\sXp}{\mathbf{X}_\text{p}}
\newcommand{\sYp}{\mathbf{Y}_{\!\text{p}}}
\newcommand{\sHhat}{\hat{\sH}}
\newcommand{\sNp}{N_\text{p}}
\newcommand{\sV}{\mathbf{V}}
\newcommand{\ssighat}{\hat{\sigma}_\swi^2}

\newcommand{\sChat}{\hat{\sC}}					
\newcommand{\sNbins}{K}							
\newcommand{\sbini}{k}							
\newcommand{\spdom}{\phi}						
\newcommand{\shistkb}{\Xi_{\sj,\sbini}^\sbit}			
\newcommand{\shistkbt}{\Xi_{\sj,\sbini}^{\sbit'}}		

\graphicspath{{xfig/}{plots/}}

\begin{document}
%
\title{Performance Assessment of MIMO-BICM\\[-4mm] Demodulators based on System Capacity}

\author{Peter~Fertl,~\IEEEmembership{Student Member,~IEEE,}
        Joakim~Jald{\'e}n,~\IEEEmembership{Member,~IEEE,}
        and~Gerald~Matz*,~\IEEEmembership{Senior~Member,~IEEE}
\thanks{
{P.~Fertl is with BMW Forschung und Technik GmbH, Hanauer Str. 46, 80992 Munich, Germany (phone: +49 89 382 53408, email: peter.fertl@bmw.de).
J.~Jald\'en is with the Signal Processing Laboratory, ACCESS Linnaeus Center, KTH Royal Institute of Technology, 
Osquldas v\"ag 10, SE-100 44 Stockholm, Sweden (phone: +46 8 790 77 88, email: jalden@kth.se).
G.~Matz is with the Institute of Communications and Radio-Frequency Engineering, Vienna University of Technology, Gusshausstrasse 25/389, A-1040 Vienna, Austria (phone: +43 1 58801 38942, email: gmatz@nt.tuwien.ac.at).}}
\thanks{Part of this work has been previously presented at {IEEE} 9th Workshop on Signal Processing Advances in Wirelesss Communications (SPAWC\;2008) \cite{Fertl:2008ac}. 
This work was supported by the STREP project MASCOT (IST-026905) within the Sixth Framework Programme of the European Commission and by FWF Grant N10606 ``Information Networks.''}}

\markboth{}{}
%



\maketitle

\vspace*{-6mm}

\begin{abstract}
\NEWa{We provide a comprehensive performance 
comparison of soft-output and hard-output demodulators in the context of
non-iterative
multiple-input multiple-output  
bit-interleaved coded modulation (MIMO-BICM).}
Coded bit error rate (BER), widely used in literature for demodulator comparison, 
has the drawback of depending strongly on the error correcting code being used.
This motivates us to propose a code-independent performance measure in terms of {\em system capacity},
i.e., mutual information of the equivalent modulation channel that comprises modulator, wireless channel, and demodulator.
We present extensive numerical results for ergodic and quasi-static fading channels 
under perfect and 
imperfect channel state information. These results reveal that the performance ranking of MIMO demodulators is rate-dependent.
Furthermore, they provide new insights regarding MIMO-BICM system design, 
i.e., the choice 
of antenna configuration, symbol constellation,
and demodulator for a given target rate.
\end{abstract}
\vspace*{-3mm}

\vspace*{-3mm}
 \ifCLASSOPTIONpeerreview
 \begin{center} \bfseries EDICS Category: MSP-CAPC, MSP-CODR \end{center}
 \fi
 
 \vspace*{-4mm}
%
%
%
%
%

\section{Introduction}
\label{sec:intro}

\subsection{Background}
\IEEEPARstart{B}{it-interleaved}
coded modulation (BICM) \cite{caire98,Guillen-i-Fabregas:2008aa} has been conceived as a pragmatic approach to coded modulation. It has received a lot of attention in wireless communications due to its bandwidth and power efficiency and its robustness against fading. For single-antenna systems, BICM with Gray labeling can approach channel capacity \cite{caire98,wachsfihuber99}. These advantages have motivated extensions of BICM to
multiple-input multiple-output (MIMO) systems \cite{bigtarvit00,Stefanov:2001,mullwein02}.

In MIMO-BICM systems, the optimum demodulator is the soft-output maximum a posteriori (MAP) demodulator, which provides the channel decoder with 
log-likelihood ratios (LLRs) for the code bits.
Due to its high computational complexity, numerous alternative demodulators have been proposed in the literature. 
Applying the max-log approximation \cite{mullwein02} to the MAP demodulator reduces complexity without significant performance loss and leads to a search for data vectors minimizing a Euclidean norm.
Exact implementations of the max-log MAP detector based on sphere decoding have been presented in \cite{jalden05_asilo,jalden_tsp05,Studer:2007aa};
sphere decoder variants in which the Euclidean norm is replaced with 
the $\slinf$ norm have been proposed in \cite{Burg:2005aa,Seethaler:2010aa}.
However, the complexity of sphere decoding grows exponentially with the number of transmit antennas \cite{jalden_tsp05}.
An alternative 
demodulator that yields approximations to the true LLRs is based on semidefinite relaxation (SDR)
and has polynomial worst-case complexity \cite{wong_sp02,Steingrimsson:2003aa}.

Several demodulation schemes 
use a list of candidate data vectors to obtain approximate LLRs.
The size of the candidate list offers a trade-off between performance and complexity.
The candidate list can be generated using 
i) tree search techniques as with the list sphere decoder (LSD) \cite{hochbrink03}, 
ii) lattice reduction (LR) techniques \cite{yao02,wubb04b,Windpassinger:2003aa,Silvola:2006aa,Ponnampalam:2007aa},
or 
iii) bit flipping techniques, i.e., flipping some of the bits in the label of a data vector obtained by hard
detection, e.g.~\cite{Wang:2006aa}.


MIMO demodulators with still smaller complexity consist of a linear equalizer followed by per-layer scalar soft demodulators.
This approach has been studied using zero-forcing (ZF) equalization \cite{butler04,McKay:2005aa} and 
minimum mean-square error (MMSE) equalization \cite{Collings:2004aa,seethal_globe04}.
The soft interference canceler (SoftIC) proposed in \cite{choi_wcnc00} iteratively performs parallel MIMO interference cancelation by subtracting an interference estimate which is computed using soft symbols from the preceding iteration.

Hard-output MIMO demodulators are alternatives to soft demodulators
that provide tentative decisions for the code bits but no associated reliability information.
Among the best-known schemes here are 
maximum likelihood (ML), 
ZF, 
and MMSE demodulation \cite{rohit03} and successive interference cancelation (SIC) \cite{wolniansky98,hassibi00,wubb03}.

\subsection{Contributions}
In the context of MIMO-BICM, the performance of the MIMO demodulators listed above has mostly been
assessed 
in terms of coded bit error rate (BER) using a specific channel code. 
These BER results depend strongly on the channel code 
and hence render an impartial demodulator comparison difficult.

In this paper, we advocate an information theoretic approach for assessing the performance of (soft and hard) MIMO demodulators in the context of non-iterative\footnote{\NEW{A performance assessment of 
\emph{soft-in} soft-out demodulators in iterative BICM receivers requires a completely different approach 
and is thus beyond the scope of this paper.}} (single-shot) BICM receivers (see also \cite{Fertl:2008ac}).
Inspired by \cite{bigtarvit00}, we propose 
the mutual information between the modulator input bits and the associated MIMO demodulator output
as a {\em code-independent\/} performance measure.
This quantity can be interpreted as system capacity (maximum rate allowing for error-free information recovery) of an equivalent ``modulation'' channel that comprises
modulator and 
demodulator in addition to the physical channel.
This approach establishes a systematic framework for the assessment of 
MIMO demodulators.
We note that ZF-based and max-log demodulation have been compared in a similar spirit in \cite{McKay:2005aa}.

Using Monte Carlo simulations, this paper provides extensive performance evaluations and comparisons for the above-mentioned MIMO demodulators in terms of system capacity,
considering different system configurations in fast and quasi-static fading.
We also investigate the performance loss of the various demodulation schemes 
under imperfect channel state information (CSI).
Due to lack of space, only a part of our numerical results is shown here. Further results for other antenna configurations, symbol constellations, and bit mappings
can be found in a supporting document \cite{Fertl:2009ac}.

Our results allow for several conclusions. Most importantly, we found that no universal performance ranking of MIMO demodulators exists, i.e., the ranking depends on the information rate or, equivalently, on the signal-to-noise ratio (SNR).
As an example, soft MMSE outperforms hard ML at low rates while at high rates it is the other way around.
We also verify this surprising observation in terms of BER simulations using low-density parity-check (LDPC) codes. 
Finally, we use our numerical results to develop practical guidelines for the design of MIMO-BICM systems, i.e., which antenna configuration, symbol constellation, and demodulator to choose in order to achieve a certain rate with minimum SNR.

\subsection{Paper Organization}
The rest of this paper is organized as follows.
Section \ref{sec:system} discusses the MIMO-BICM system model and Section \ref{sec:measure} proposes system capacity as performance measure.
In Sections \ref{sec:basic} and \ref{sec:other}, we assess the system capacity achievable with the MIMO-BICM demodulators referred to above for the case of fast fading. Section \ref{sec:imperfect} analyzes the impact of imperfect channel state information (CSI) on the demodulator performance, and Section \ref{sec:slow} investigates the rate-versus-outage tradeoff of selected demodulators in quasi-static environments.
In Section \ref{sec:practic}, we summarize key observations and infer practical system design guidelines.
Finally, conclusions are provided in Section \ref{sec:conclusion}.

\begin{figure*}
\centering
\input{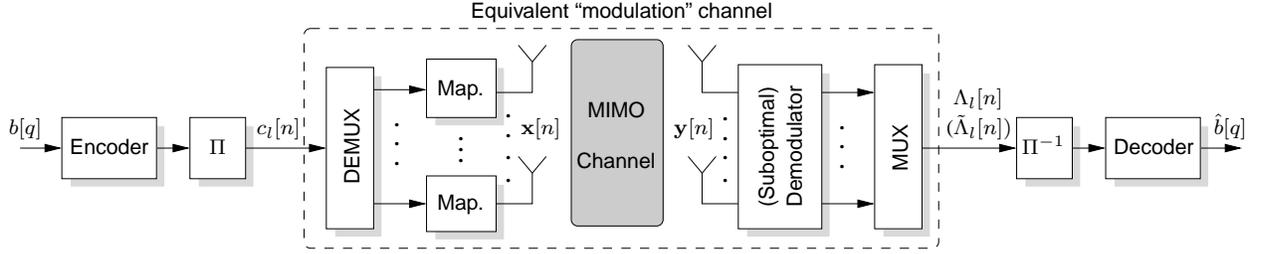}
\vspace*{-11mm}
\caption{Block diagram of a MIMO-BICM system.}
\label{fig:mimo-bicm}
\vspace*{-7mm}
\end{figure*}

\section{System Model}
\label{sec:system}

\subsection{MIMO-BICM Transmission Model}
A block diagram of our MIMO-BICM model is shown in Fig.~\ref{fig:mimo-bicm}.
The information bits $\sbit[\sq]$ are encoded using an error-correcting code and is then passed through a
{\em bitwise\/} interleaver
$\Pi$. The interleaved code bits 
are demultiplexed into $\sMt$ antenna streams (``layers'').
In each layer $\sk=1,\dots,\sMt$, groups of $\sm$ code bits
$\scik[\sn]$, $\si=1,\dots,\sm$, ($\sn$ denotes symbol time)
are mapped via a one-to-one function $\smu(\cdot)$ to
(complex) data symbols $\sxi_\sk[\sn]$ from a symbol alphabet $\sA$ of size $\sM\!=\!2^\sm$.
Specifically, $\sxi_\sk[\sn] = \smu(\scbit_\sk^{(1)}[\sn],\dots,\scbit_\sk^{(\sm)}[\sn])$,
where $\big\{\scbit_\sk^{(1)}[\sn],\dots,\scbit_\sk^{(\sm)}[\sn]\big\} = \smu^{-1}(\sxi_\sk[\sn])$ is referred to as the bit label of $\sxi_\sk[\sn]$.
The transmit vector 
is given by
\footnote{The superscripts $\strans$ and $\sherm$ denote transposition and Hermitian transposition, respectively.
Furthermore, $\shA\!\define\!\sA\!\times\!\dots\!\times\!\sA$ is the $\sMt$-fold Cartesian product of
$\sA$, $\sE\{\cdot\}$ denotes expectation, and $\|\cdot\|$ is the $\sltwo$ (Euclidean) norm.} 
$\sx[\sn] \define (\sxi_1[\sn] \dots \sxi_{\sMt}[\sn])\strans \in \shA$
and satisfies the power constraint $\sE\{\|\sx[\sn]\|^2\}=\sPt$.
It carries $\slambda\!=\!\sm\sMt$ interleaved code bits
$\scj[\sn]$, $\sj\!=\!1,\dots,\slambda$, with $\scik[\sn] = \scbit_{(\sk-1)\sm+\si}[\sn]$.
We will for simplicity write $\sx[\sn] = \smu(\scbit_1[\sn],\dots,\scbit_{\slambda}[\sn])$
and
$\svc[\sn] \!\define\! (\scbit_1[\sn] \dots \scbit_{\slambda}[\sn])\strans
=\smu^{-1}(\sx[\sn])$
as shorthand for the mapping $\sx[\sn] = \big( \smu(\scbit_1^{(1)}[\sn],\dots,\scbit_1^{(\sm)}[\sn]) \,\dots\,  \smu(\scbit_{\sMt}^{(1)}[\sn],\dots,\scbit_{\sMt}^{(\sm)}[\sn]) \big)\strans$ and its inverse.



Assuming flat fading, the receive vector $\sr[\sn]\!\define\!(\sri_1[\sn] \dots \sri_{\sMr}[\sn])\strans$
($\sMr$ denotes the number of receive antennas)
is given by
\begin{equation}
\sr[\sn] = \sH[\sn]\hspace*{.3mm} \sx[\sn]+\sw[\sn]\,, \qquad \sn=1,\dots,\sN\,, \label{eq:system-model}
\end{equation}
where $\sH[\sn]$ is the $\sMr\times\sMt$ channel matrix, and
$\sw[\sn]\define(\swi_1[\sn] \dots \swi_{\sMr}[\sn])\strans$ is a noise vector
with independent identically distributed (i.i.d.) circularly symmetric complex Gaussian elements with zero mean and variance $\ssigw$.
In most of what follows, we will omit the time index $\sn$ for convenience.


At the receiver, the optimum demodulator uses the received vector $\sr$ and the channel matrix $\sH$
to calculate LLRs $\sLLRj$ for all code bits $\scj$, $\sj\!=\!1,\dots,\slambda$, carried by $\sx$.
In practice, the use of suboptimal demodulators or of a channel estimate $\hat{\sH}$ will result in approximate LLRs $\shLLRj$.
The LLRs are 
passed through the deinterleaver $\Pi^{-1}$ and then on to the channel decoder that delivers the detected bits $\sbithat[\sq]$.

\subsection{Optimum Soft MAP Demodulation}
Assuming i.i.d.\ uniform code bits (as guaranteed, e.g., by an ideal interleaver),
the optimum soft MAP demodulator calculates the exact LLR for $\scj$ based on $(\sr,\sH)$ according to \cite{mullwein02}
\begin{align}
\sLLRj&\define \log 
\frac{\spmf(\scj\!=\!1|\sr,\sH)}{\spmf(\scj\!=\!0|\sr,\sH)}
=\log 
\frac{%
\sum\limits_{\sx\in\shAo} \exp\left(-\frac{\|\sr-\sH\sx\|^2}{\ssigw}\right)%
}{%
\sum\limits_{\sx\in\shAz} \exp\left(-\frac{\|\sr-\sH\sx\|^2}{\ssigw}\right)%
} 
\,. \label{eq:exact-LLR}
\end{align}
Here, $\spmf(\scj|\sr,\sH)$ is the probability mass function (pmf) of the code bits conditioned on $\sr$ and $\sH$, $\shAo$ and $\shAz$ denote the complementary sets of transmit vectors for which $\scj\!=\!1$ and $\scj\!=\!0$, respectively (note that $\shA=\shAo \cup \shAz$). 
Unfortunately, computation of \eqref{eq:exact-LLR} has complexity $\sO(\sM^{\sMt})\!=\!\sO(2^{\slambda})$, i.e.,
exponential in the number of transmit antennas.
For this reason, several suboptimal demodulators have been proposed which promise near-optimal performance while requiring a lower computational complexity. The aim of this work is to provide a fair performance comparison of these demodulators.

\section{System Capacity}
\label{sec:measure}
In order for the information rates discussed below to have interpretations as ergodic capacities, we consider a fast fading scenario where 
the channel $\sH[\sn]$ is a stationary, finite-memory process. 
We recall that the ergodic capacity with Gaussian inputs is given by\NEW{\cite{telatar_ett99}}
\begin{equation}\label{eq:gausscap}
\sC_\text{G}=\sE_\sH\Big\{\log_2 \det \Big(\seye +\frac{\sPt}{\sMt\ssigw}\,\sH\sH\sherm\Big)\Big\}
\end{equation}
(here, $\seye$ denotes the identity matrix).
The non-ergodic regime (slow fading) is discussed in Section \ref{subsec:slow-fading}. 

\subsection{Capacity of MIMO Coded Modulation}
In a coded modulation (CM) system with equally likely transmit vectors $\sx\in\shA$ and no CSI at the transmitter,
the average mutual information in bits per channel use (bpcu) is given by
(cf.\ \cite{bigtarvit00})
\begin{equation}
\sC_\text{CM} \hspace*{-.3mm} \define\hspace*{-.3mm}  \sI(\sx;\sr|\sH) \hspace*{-.3mm} = \hspace*{-.3mm}
\slambda \hspace*{.3mm} -\hspace*{.3mm}  \sE_{\sx,\sr,\sH}\!\left\{\!\log_2\! \frac{\sum\limits_{\sx'\in\shA} \sprob(\sr|\sx',\sH)}{\sprob(\sr|\sx,\sH)}\!\right\}\!. \label{eq:CM-eval}
\end{equation}
Here, 
we used
the conditional
probability density function (pdf) (cf.\ \eqref{eq:system-model})  
\begin{equation}
\sprob(\sr|\sx,\sH)=\frac{1}{(\pi\ssigw)^{\sMr}}\exp\left(-\frac{\|\sr-\sH\sx\|^2}{\ssigw}\right). \label{eq:prob-dens}
\end{equation}
In the following, we will refer to $\sC_\text{CM}$ as {\em CM capacity} \cite{caire98}
(sometimes, $\sC_\text{CM}$ is alternatively termed constellation-constrained capacity).
It is seen from \eqref{eq:CM-eval} that $\sC_\text{CM} \le \slambda$
; in fact, the last term in \eqref{eq:CM-eval} may be interpreted as a penalty term resulting from the noise and MIMO interference.

Using the fact that the mapping between the symbol vector $\sx$ and the associated bit label $\{\scbit_1,\dots,\scbit_{\slambda}\}$ is one-to-one and applying the chain rule for mutual information \cite[page~$24$]{cover91} to \eqref{eq:CM-eval}
leads to 
\begin{align}
\sC_\text{CM}
& = \sI(\scbit_{1},\dots,\scbit_{\slambda};\sr|\sH) \label{eq:chain} 
=\sum\limits_{\sj=1}^{\slambda} \sI(\scbit_{\sj};\sr|\scbit_{1},\dots,\scbit_{\sj-1},\sH)\\ & = \sum\limits_{\sj=1}^{\slambda} \sent(\scbit_{\sj}|\sH) - \sent(\scbit_{\sj}|\sr,\scbit_{1},\dots,\scbit_{\sj-1},\sH); \notag
\end{align}
here, $\sent(\cdot)$ denotes the entropy function.
The single-antenna equivalent of \eqref{eq:chain} served as a motivation for multilevel coding and multistage decoding, which
can indeed achieve CM capacity \cite{wachsfihuber99}. 
\NEW{
Multilevel coding for multiple antenna systems has been considered in \cite{Lampe:2004aa}.}

\subsection{Capacity of MIMO-BICM}
\NEW{In the following, we assume an {\em ideal, infinite-length} bit interleaver\footnote{In practice, this means that the interleaver needs to be much longer than the 
codewords transmitted over the channel.} which allows us to treat the BICM system as a set of $\slambda$ independent parallel memoryless binary-input channels as in \cite[Section III.A]{caire98}. 
}
Using the assumption of i.i.d.\ uniform code bits%
, the maximum rate achievable with BICM 
is given by (cf.\ \cite{bigtarvit00})
\begin{align}
\sC_\text{BICM} &\triangleq \sum_{\sj=1}^{\slambda} \sI(\scj;\sr|\sH)= \sum\limits_{\sj=1}^{\slambda} \sent(\scj|\sH) - \sent(\scj|\sr,\sH)
\label{eq:BICM} \\
&= \slambda \hspace*{.1mm} - \hspace*{.1mm}\sum_{\sj=1}^{\slambda}\, \sE_{\sx,\sr,\sH}\!\left\{ \log_2\! \frac{\sum\limits_{\sx'\in\shA} \!\sprob(\sr|\sx',\sH)}{\sum\limits_{\sx'\in \mathcal{X}_\sj^{\scj} }\sprob(\sr|\sx',\sH)}\right\}\,, \notag
\end{align}
where\footnote{By $(\mathbf{x})_\sk$ and $({\bf X})_{k,l}$ we respectively denote the $k$th element of the vector $\sx$ and the element in row $k$ and column $l$ of the matrix ${\bf X}$.} 
$\scj=(\svc)_\sj=(\smu^{-1}(\sx))_\sj$ denotes the $\sj$th bit in the label of $\sx$. 
\NEW{Since conditioning reduces entropy \cite[page~$29$]{cover91}, a comparison of 
\eqref{eq:chain} 
and 
\eqref{eq:BICM} 
reveals that \cite{Lampe:2004aa}}
\begin{equation}
\sC_\text{BICM} 
\,\le\,
\sC_\text{CM}\,. 
\notag
\end{equation}
The gap $\sC_\text{CM}-\sC_\text{BICM}$ increases with $\sM$ and $\sMt$ and depends strongly on the symbol labeling \cite{mullwein02}.
For single-antenna BICM systems with Gray labeling, this gap has been shown 
to be negligible \cite{caire98,wachsfihuber99}; however, for MIMO systems (see Section \ref{sec:basic}) 
\NEW{and at low SNRs in the wideband regime \cite{Guillen-i-Fabregas:2008aa}
it can be significant. The capacity loss can be attributed to the fact that the BICM receiver neglects the dependencies between the transmitted code bits.}
\NEW{Under the unrealistic assumption of perfectly known channel SNR, multilevel coding with multistage decoding can in principle avoid such a capacity loss but suffers from error propagation \cite{wachsfihuber99,Lampe:2004aa}. BICM does not require the channel SNR at the transmitter and can be considered more robust.   
A hybrid version of CM and BICM whose complexity and performance is between the two
was presented in \cite{Lampe:2004aa}. 
Furthermore, augmenting BICM with space-time mappings can be beneficial (cf.~\cite{Hong:2001aa,Lampe:2004aa})
but is not considered here due to space limitations.}

It can be shown that the log-likelihood ratio $\sLLRj$ in \eqref{eq:exact-LLR} is a sufficient statistic \cite{van-Dijk:2003aa} for $\scj$ given $\sr$ and $\sH$. 
Therefore, \eqref{eq:BICM} can be rewritten as
\begin{equation} \label{eq:C_BICM_LLR}
\sC_\text{BICM} 
= \sum_{\sj=1}^{\slambda}\sI(\scj;\sLLRj)\,. 
\end{equation} 
Hence, $\sC_\text{BICM}$ can be interpreted as the capacity of an equivalent channel with inputs $\scj$ and outputs $\sLLRj$ (cf.~Fig.~\ref{fig:mimo-bicm}). This channel is characterized by the conditional pdf
$\prod\limits_\sj\sprob(\sLLRj|\scj)$, which usually is hard to obtain analytically, however.

\subsection{System Capacity and Demodulator Performance}
Motivated by the interpretation of $\sC_\text{BICM}$ as the system capacity of BICM using the optimum MAP demodulator, we propose to measure the performance of sub-optimal MIMO-BICM demodulators via the system capacity of the associated equivalent ``modulation'' channel with {\em binary} inputs $\scj$ and 
the approximate LLRs
$\shLLRj$ as {\em continuous} outputs (cf.\ Fig.~\ref{fig:mimo-bicm}).
This channel is described by the conditional pdf $\prod\limits_\sj\sprob(\shLLRj|\scj)$.
Its system capacity is defined as the mutual information between $\scj$ and $\shLLRj$,
which can be shown to equal
\begin{align}
\label{eq:measure-exact}
\sR
\;\triangleq \;\sum_{\sj=1}^{\slambda} \sI(\scj;\shLLRj)
\;=\;\slambda \hspace*{.3mm} - \hspace*{.3mm} \sum_{\sj=1}^{\slambda} \sum_{b=0}^1 \int_{-\infty}^{\infty}
\frac{1}{2}\sprob(\shLLRj|\scj\!=\!b) \log_2 \!\frac{\sprob(\shLLRj)}{\frac{1}{2}\sprob(\shLLRj|\scj\!=\!b)} d\shLLRj \,, 
\end{align}%
where
$\sprob(\shLLRj)\!=\!
\big[\sprob(\shLLRj|\scj\!=\!0)+\sprob(\shLLRj|\scj\!=\!1)\big]/2$.
We emphasize that the system capacity $\sR$ provides a performance measure for MIMO (soft) demodulators that is independent of the outer channel code.
In fact, it has an intuitive operational interpretation as the highest rate
achievable (in the sense of asymptotically vanishing error probability)
in a BICM system \NEW{with independent parallel channels (
assumption of an ideal infinite-length interleaver, cf.\ \cite[Section III.A]{caire98})}, using the specific demodulator which produces $\shLLRj$. 
Since $\shLLRj$ is derived from $\sr$ and $\sH$,
the data processing inequality \cite[page $34$]{cover91} implies that $\sR \leq \sC_\text{BICM}$ with equality if
$\shLLRj$ is a one-to-one function of $\sLLRj$.
The performance of a soft demodulator can thus be measured in terms of the gap  $\sC_\text{BICM} - \sR$.
Of course, the information theoretic performance measure in \eqref{eq:measure-exact} does not take into account complexity issues
and it has to be expected that a reduction of the gap $\sC_\text{BICM} - \sR$ in general can only be achieved at the expense of
increasing computational complexity.

We caution the reader that the rates in \eqref{eq:C_BICM_LLR} and \eqref{eq:measure-exact}
are sums of mutual informations for the individual code bits $c_1,\dots,c_{\slambda}$ carried by one symbol vector. 
Indeed, the pdfs $\sprob(\sLLRj|\scj)$ and $\sprob(\shLLRj|\scj)$ in general depend on the code bit position $\sj$, even though for certain systems (e.g.\ 4-QAM modulation) the code bit protection and LLR statistics are independent of the bit position $\sj$
for reasons of symmetry.
Achieving \eqref{eq:C_BICM_LLR} and \eqref{eq:measure-exact} thus requires channel encoders and decoders that take the bit position into account.
When the channel code fails to use this information, the rate loss is small
provided that the mapping protects different code bits $\scj$ roughly equally against noise and interference.

\subsection{Non-Ergodic Channels} \label{subsec:slow-fading}
In the case of quasi-static or slow fading \cite{tsevis05},
the channel $\sH$ is random but constant over time, i.e., each codeword can extend over only one channel realization.
Here, the ergodic capacity of the modulation channel is no longer operationally meaningful
\cite{tsevis05,Lapidoth:1998}. Instead we consider the \emph{outage probability}
\begin{equation} \label{eq:OUTAGE_PROBABILITY}
\sPout(\sRbar) \triangleq \sP\{\sRH \!<\! \sRbar\} ,
\end{equation}
where $\sRH$ is a random variable defined as
\begin{equation*} 
\sRH \triangleq \sum_{\sj=1}^{\slambda} \sI_{\sH}(\scj;\shLLRj).
\end{equation*}
Here, $\sI_{\sH}(\scj;\shLLRj)$ denotes the conditional mutual information, which is evaluated with $\sprob(\shLLRj|\scj,\sH)$ in place of $\sprob(\shLLRj|\scj)$ (cf.\ \eqref{eq:measure-exact}). Note that the ergodic system capacity $\sR$ in \eqref{eq:measure-exact} equals $\sR = \sE_{\sH} \{ \sRH \}$. The outage probability $\sPout(\sRbar)$ can be interpreted
as the smallest codeword error probability achievable at rate $\sRbar$ \cite{Lapidoth:1998}. A closely related concept is the \emph{$\epsilon$-capacity} of the equivalent modulation channel, 
defined as
\begin{equation} \label{eq:EPSILON_CAPACITY}
\sReps \triangleq \sup \; \{ \sRbar \; | \; \sP\{\sRH \! < \! \sRbar\} < \epsilon \}\,.
\end{equation}
The $\epsilon$-capacity may be interpreted as the maximum rate for which a codeword error probability
less than $\epsilon$ can be achieved. Rates smaller than $\sReps$ are referred to as $\epsilon$-achievable rates \cite{Lapidoth:1998}. 
If $\sPout(\sRbar)$ is a continuous and increasing function of $\sRbar$ (which is usually the case in practice),
it holds that $\sPout(\sReps)=\epsilon$.



\NEW{
\subsection{Generalized Mutual Information}
The operational interpretation of our performance measure as the largest achievable rate for a BICM system using a given demodulator requires the assumption of an ideal infinite-length interleaver. 
With a finite-length interleaver, the parallel channels (i.e., the different bits in a given symbol vector) are not independent in general;
here, achievable rates can be characterized in terms of the {\em generalized mutual information (GMI)} which is obtained by treating the BICM receiver as a mismatched decoder \cite{Merhav94,Guillen-i-Fabregas:2008aa,Martinez:2008aa}. For the case of optimum soft MAP demodulation (cf.~\eqref{eq:exact-LLR}), 
the BICM capacity using the independent parallel-channel model coincides with the GMI 
\cite{Martinez:2008aa}. 
%
We recently provided a non-straightforward extension of this result by showing that the GMI of a BICM system with suboptimal demodulators augmented with \emph{scalar LLR correction} (see Section \ref{ssec:ber}) coincides with the system capacity in \eqref{eq:measure-exact} obtained for the parallel-channel model \cite{Jalden:2009ab}.\footnote{We note that 
the LLR correction leaves the mutual information which underlies system capacity unchanged.}
Scalar LLR correction has been used previously 
to provide the binary decoder with accurate reliability information \cite{van-Dijk:2003aa,Burg:2006aa,Schwandter:2008aa,Novak:2009aa,Studer:2009aa}. 
The GMI of a BICM system with finite interleaver and LLR-corrected suboptimal demodulators can thus efficiently be computed by evaluating \eqref{eq:measure-exact} \cite{Jalden:2009ab}; 
this provides additional justification for the use of \eqref{eq:measure-exact} as a code-independent performance measure for approximate demodulators. We note that a GMI-based analysis of BICM with mismatched decoding metrics
that generalizes our work in \cite{Jalden:2009ab} has recently been presented in \cite{lampe10}.}

\section{Baseline MIMO-BICM Demodulators}
\label{sec:basic}
In this section, we first review max-log and hard ML demodulation as well as linear MIMO demodulators and then we provide results illustrating their performance in terms of system capacities. These demodulators serve as baseline systems for later demodulator performance comparisons in Section \ref{sec:other}. We note that max-log and hard ML MIMO demodulators have the highest complexity among all soft and hard demodulation schemes, respectively, whereas linear MIMO demodulators are most efficient computationally.
\NEW{Due to space limitations, we only state the complexity order of each demodulator in the following 
and we give references that provide more detailed complexity analyses.
}

\subsection{Max-Log and Hard ML Demodulator}
Applying the max-log approximation to \eqref{eq:exact-LLR} simplifies the LLR computation to 
a minimum distance problem and results in the approximate LLRs \cite{mullwein02}
\begin{equation}
\shLLRj = 
\frac{1}{\ssigw}\!\bigg[ {\underset{\sx\in\shAz}{\min} \|\sr-\sH\sx\|^2}\,-{\underset{\sx\in\shAo}{\min} \|\sr-\sH\sx\|^2}\bigg]. \label{eq:max-log}
\end{equation}
This expression can be implemented easier than \eqref{eq:exact-LLR}
since it avoids the logarithm and exponential functions.
However, computation of $\shLLRj$ in \eqref{eq:max-log} still requires two searches over sets of size $\sM^{\sMt}\!/2=2^{\slambda-1}$.
Sphere decoder implementations of \eqref{eq:max-log} are presented in \cite{jalden05_asilo,Studer:2007aa}.

Hard vector ML demodulation amounts to the minimum distance problem
\begin{equation}
\shxml = \underset{\sx\in\shA}{\arg\;\min}\, \|\sr-\sH\sx\|^2. \label{eq:hardML}
\end{equation}
\NEW{This optimization problem can be solved by exhaustive search or using a sphere decoder;
in both cases, the computational complexity 
scales exponentially with the number of transmit antennas.}
The detected code bits $\shcj$ corresponding to \eqref{eq:hardML} are obtained via the one-to-one mapping between code bits and symbol vectors, i.e., $\svhc=(\shcbit_1 \dots \shcbit_{\slambda})\strans = \smu^{-1}(\shxml)$.
It can be shown that the code bits $\shcj$ obtained by the hard ML detector
correspond to the sign of the corresponding max-log LLRs in \eqref{eq:max-log}, i.e.,
$\shcj = \su(\shLLRj)$ where $\su(\cdot)$ denotes the unit step function.
When it comes to computing the system capacity with hard-output demodulators,
the only difference to soft-output demodulation is the {\em discrete} nature of the outputs $\shcj$ of the equivalent ``modulation'' channel, which here becomes a binary channel.
Consequently, the integral over $\shLLRj$ in \eqref{eq:measure-exact} is replaced with a summation over $\shcj\in\{0,1\}$.

\subsection{Linear Demodulators}
\label{ssec:lin_demod}
In the following, $\sLLRik$ is the LLR corresponding to $\scik$ (the $\si$th bit in the bit label of the $\sk$th symbol $\sxi_\sk$).
Soft demodulators with extremely low complexity can be obtained by using a linear (ZF or MMSE) equalizer followed by {\em per-layer} max-log LLR calculation according to
\begin{equation}
\shLLRik =
\frac{1}{\sprefac} \bigg[ \,\underset{\sxi\in\sA_\si^0}{\min} |\shxi_\sk\!-\!\sxi|^2 - \underset{\sxi\in\sA_\si^1}{\min} |\shxi_\sk\!-\!\sxi|^2 \bigg]
,\qquad \si=1,\dots,\sm,\;\; \sk=1,\dots,\sMt\,. \label{eq:linear-LLR}
\end{equation}
Here, $\sA_\si^b\!\subset\!\sA$ denotes the set of (scalar) symbols whose bit label at position $\si$ equals $b$, $\shxi_\sk$ is an estimate of the symbol in layer $\sk$ provided by the equalizer, and $\sprefac$ is an equalizer-specific weight (see below).
We emphasize that calculating LLRs separately for each layer results in a significant complexity reduction.
In fact, 
calculating the symbol estimates $\shxi_\sk$ using a ZF or MMSE equalizer requires $\sO(\sMr\sMt^2)$ operations;
furthermore, the complexity of evaluating \eqref{eq:linear-LLR} for all code bits
scales as $\sO(\sMr\sMt 2^\sm)=\sO(\sMr\sMt \sM)$, i.e., linearly in the number of antennas{}\NEW{\cite{Collings:2004aa,seethal_globe04}}.

Equalization-based hard bit decisions $\shcik $ can be obtained by quantization of the equalizer output $\shxi_k$ with respect to $\sA$ (denoted by $\sQuant(\cdot)$),
followed by the demapping, i.e.,
$
\big( \shcbit_\sk^{(1)}\dots\, \shcbit_\sk^{(\sm)}\big)\strans
=\smu^{-1}\big( \sQuant(\shxi_\sk) \big)$.
Again, the detected code bits correspond to the sign of the LLRs, i.e., $\shcik = \su(\shLLRik)$.

\vspace{1mm}

\subsubsection{ZF-based Demodulator \cite{butler04,McKay:2005aa}}
Here, the first stage consists of ZF equalization, i.e.,
\begin{equation}
\syzf=(\sH\sherm\sH)^{-1}\sH\sherm\sr = \sx + \sv \,, \label{eq:yzf}
\end{equation}
where the post-equalization noise vector $\sv$ has correlation matrix 
\begin{equation}
\sRw = \sE\{ \sv\sv\sherm \} =
\ssigw\, (\sH\sherm\sH)^{-1}. \label{eq:zf-cov}
\end{equation}
Subsequently, approximate bit LLRs are obtained according to \eqref{eq:linear-LLR}
with symbol estimate $\shxi_\sk = (\syzf)_\sk$ and weight factor $\sprefac = (\sRw)_{\sk,\sk}$.

\vspace{1mm}

\subsubsection{MMSE-based Demodulator \cite{seethal_globe04,Collings:2004aa}}
Here, the first stage is an MMSE equalizer that can be written as (cf.\ \eqref{eq:yzf} and \eqref{eq:zf-cov})
\begin{equation}
\symmse=\sW\syzf\,, \qquad\text{with}\;\;\;
\sW = \left(\seye + \frac{\sMt}{\sPt}\sRw\right)^{-1}. \label{eq:wiener}
\end{equation}
Approximate LLRs are then calculated according to \eqref{eq:linear-LLR} with 
\begin{align*}
\shxi_\sk = \frac{(\symmse)_\sk}{\sWkk}
\quad\text{and} \quad
\sprefac =  \frac{\sPt}{\sMt}\frac{1-\sWkk}{\sWkk},
\end{align*}
where $\sWkk=(\sW)_{\sk,\sk}$.
\NEW{Here, $\shxi_\sk$ denotes the output of the {\em unbiased} MMSE equalizer, which is preferable to
a biased MMSE equalizer for non-constant modulus modulation schemes such as $16$-QAM and $64$-QAM \cite{cioffi95};
in the remainder we will thus restrict to unbiased soft and hard MMSE demodulators.}

\begin{figure*}
\centering
\subfigure[]{\includegraphics[scale=0.365]{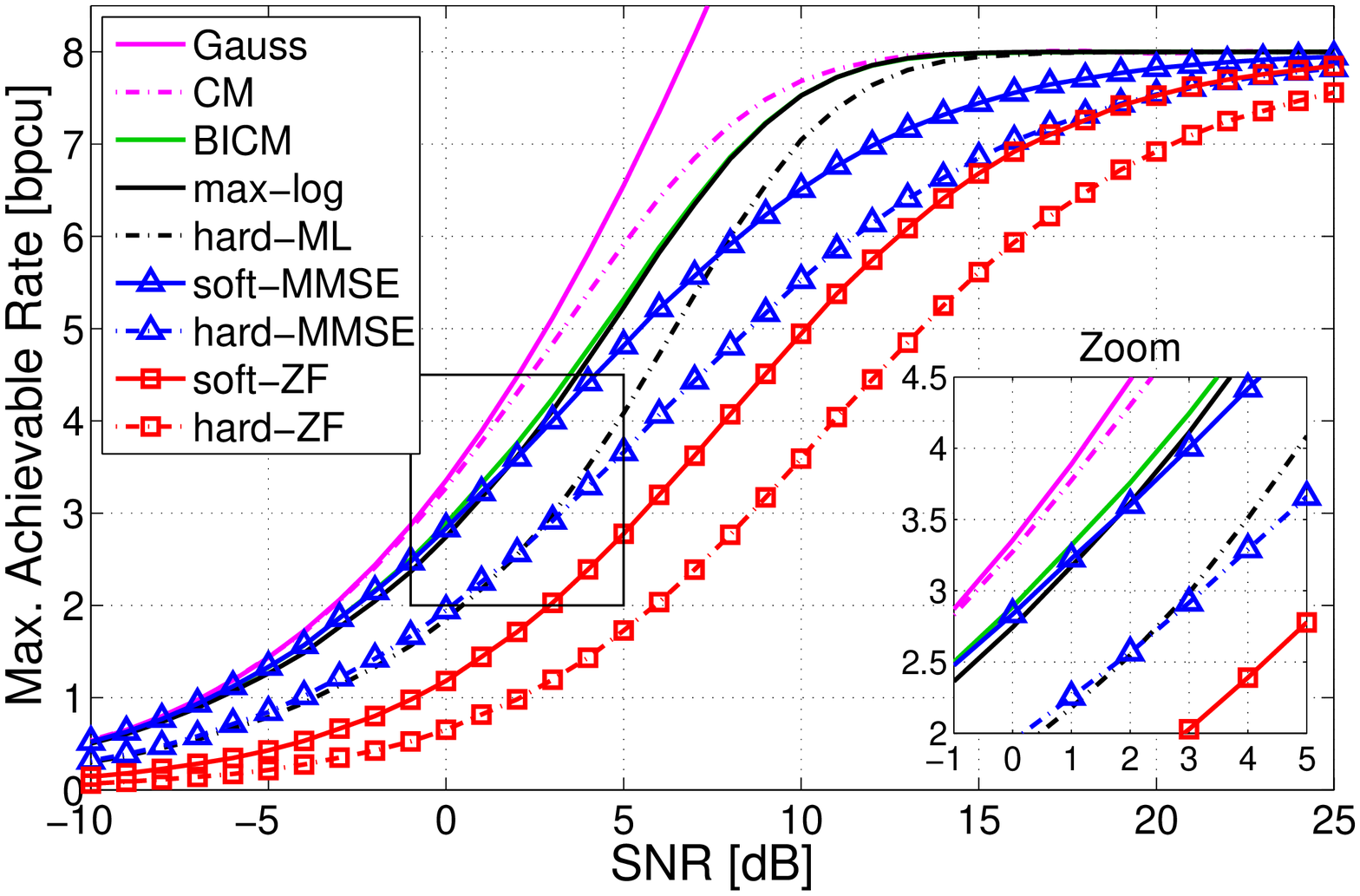}}
~\hfill~
\subfigure[]{\includegraphics[scale=0.365]{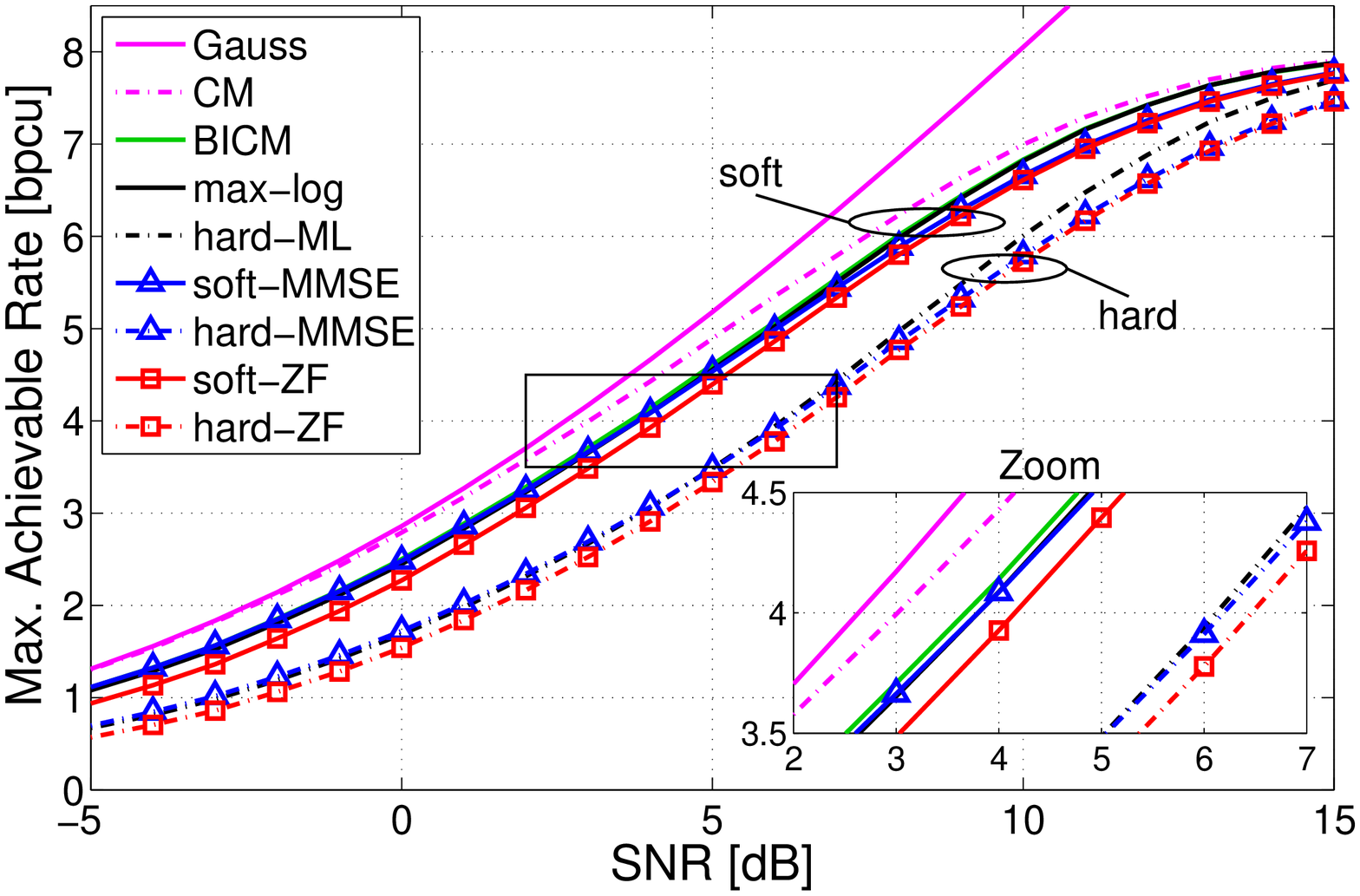}}
\vspace*{-9mm}
\caption{\label{fig:basic_gray}Numerical capacity results for (a) a $4\!\times\!4$ MIMO system with $4$-QAM, and
(b) a $2\!\times\!4$ MIMO system with $16$-QAM (in both cases with Gray labeling).}
\vspace*{-7mm}
\end{figure*}

\subsection{Capacity Results}
We next compare the performance of the above baseline demodulators (i.e., max-log, hard ML, MMSE and ZF) in terms of their maximum achievable rate $\sR$ (ergodic system capacity, see \eqref{eq:measure-exact}). In addition, the CM capacity $C_\text{CM}$ in \eqref{eq:CM-eval}, the MIMO-BICM capacity $C_\text{BICM}$ in \eqref{eq:BICM} \NEW{(corresponding to the system capacity of the optimum soft MAP demodulator in \eqref{eq:exact-LLR})},
and the Gaussian input channel capacity $C_\text{G}$ in \eqref{eq:gausscap} are shown as benchmarks.
Throughout the paper, all capacity results have been obtained for spatially i.i.d.\ Rayleigh fading, with all fading coefficients normalized to unit variance.

The pdfs required for evaluating 
\eqref{eq:measure-exact}
are generally hard to obtain in closed form. Thus, we measured these pdfs using
Monte Carlo simulations and then evaluated all integrals numerically. 
Based on the results in \cite{Paninski:2003aa}, we numerically optimized the binning (used to measure the pdfs) in order to reduce the bias and variance of the mutual information estimates \NEW{(see Appendix \ref{app:estimate-MI} for more details)}. 
The capacity results (obtained with $10^5$ fading realizations) are shown in Fig.~\ref{fig:basic_gray} in bits per channel use versus SNR $\sSNR\define{\sPt}/{\ssigw}$.
In the following, in some of the plots we show insets that provide zooms of the capacity curves around a rate of $\slambda/2$\,bpcu.


Fig.~\ref{fig:basic_gray}(a) pertains to the case of $\sMt\times\sMr=4\times4$ MIMO 
with Gray-labeled $4$-QAM (here, $\slambda=8$).
At a target rate of $4$\,bpcu, 
the SNR required for CM and Gaussian capacity is virtually the same, whereas that for BICM is larger by about $1.3\,$dB.
The SNR penalty of using max-log demodulation instead of soft MAP is about $0.3$\,dB. Furthermore, hard ML demodulation requires a $2.1$\,dB higher SNR to achieve this rate than max-log demodulation; for soft and hard MMSE demodulation the SNR gaps to max-log are $0.2$\,dB and $3.1$\,dB, respectively, while for soft and hard ZF demodulation they respectively equal $5.1$\,dB and $8.1$\,dB.
\NEW{An interesting observation in this scenario is the fact that at low rates, soft and hard MMSE demodulation slightly outperform max-log and hard ML demodulation, respectively, whereas at high rates MMSE demodulation degrades to ZF performance.}
\NEW{Hard MMSE demodulation can outperform hard ML demodulation since the latter minimizes the vector symbol error 
probability whereas our system capacity is defined on the bit level.
}
Surprisingly, at low rates soft MMSE essentially coincides with BICM capacity.
Moreover, soft MMSE demodulation outperforms hard ML demodulation at low-to-medium rates whereas at high rates it is the other way around (the cross-over can be seen at about $5.8\,$bpcu).
These observations reveal the somewhat unexpected fact that the demodulator performance ranking is not universal but depends on the target rate (or equivalently, the target SNR), even if the number of antennas, the symbol constellation, and the labeling are fixed.
Similar observations apply to $16$-QAM instead of $4$-QAM and to set-partitioning labeling instead of Gray labeling (see \cite{Fertl:2009ac}). Apart from a general shift of all curves to higher SNRs, 
the larger constellation and/or the different labeling strategy causes an increase of the gap between CM capacity and BICM capacity.
The gaps between hard ML, hard MMSE, and soft ZF demodulation are significantly smaller, though, in this case
(soft ZF outperforms hard MMSE for rates above $6.2$\,bpcu and approaches hard ML for rates around $6$\,bpcu). When decreasing the antenna configuration to a $2\times 2$ system, we observed that soft ZF outperforms hard ML demodulation for low-to-medium rates, e.g., by about $1.7$\,dB at $4$\,bpcu with $16$-QAM \cite{Fertl:2009ac}. 

The situation changes for the case of a $2\times4$ MIMO system with Gray-labeled $16$-QAM (again $\slambda=8$), shown in Fig.~\ref{fig:basic_gray}(b).
The increased SNR gap between CM and BICM capacity implied by the larger constellation
is compensated by having more receive than transmit antennas (this agrees with observations in \cite{mullwein02}).
In addition, the performance differences between the individual demodulators are significantly reduced, revealing an essential distinction being between soft and hard demodulators.
Having $\sMr>\sMt$ helps the linear demodulators approach their non-linear counterparts even at larger rates, i.e., soft ZF/MMSE perform close to max-log and hard ZF/MMSE perform close to hard ML, with an SNR gap of about $2.3$\,dB between hard and soft demodulators.
Note that in this scenario soft MMSE and soft ZF both outperform hard ML demodulation at all rates.

\subsection{BER Performance}
\label{ssec:ber}
Even though we advocate a demodulator comparison in terms of system capacity, the cross-over of some of the capacity curves prompts a verification in terms of the BER of soft and hard MMSE demodulation as well as max-log and hard ML demodulation.
We consider a $4\times4$ MIMO-BICM system with Gray-labeled $4$-QAM in conjunction with irregular LDPC codes\footnote{The LDPC code design was performed using the web-tool at {\tt http://lthcwww.epfl.ch/research/ldpcopt}.} \cite{richurb01} of block length $64000$.
For the case of soft demodulation, the LDPC codes were designed for an additive white Gaussian noise (AWGN) channel whereas for the case of hard demodulators the design was for a binary symmetric channel. 
At the receiver, message-passing LDPC decoding \cite{richurb01} was performed. 
In the case of hard demodulation, the message-passing decoder was provided with the LLRs
\begin{equation} \label{eq:hardllrs}
\shLLRj 
= (2\shcj\!-\!1)\,\log \frac{1\!-\!\spcross}{\spcross},
\end{equation}
where $\spcross=\sP\{\shcj\ne\scj|\scj\}$, the cross-over probability of the equivalent binary channel, was determined via Monte Carlo simulations.
\NEW{With the soft demodulators, we performed
an LLR correction via a lookup table as in \cite{Studer:2009aa}. 
Using LLR correction for soft demodulators and \eqref{eq:hardllrs} for hard output demodulators is 
critical in order to provide the channel decoder 
with accurate reliability information \cite{van-Dijk:2003aa,Burg:2006aa,Schwandter:2008aa,Novak:2009aa,Studer:2009aa,Jalden:2009ab}.}

\begin{figure*}[t]
\centering
\subfigure[]{\includegraphics[scale=0.363]{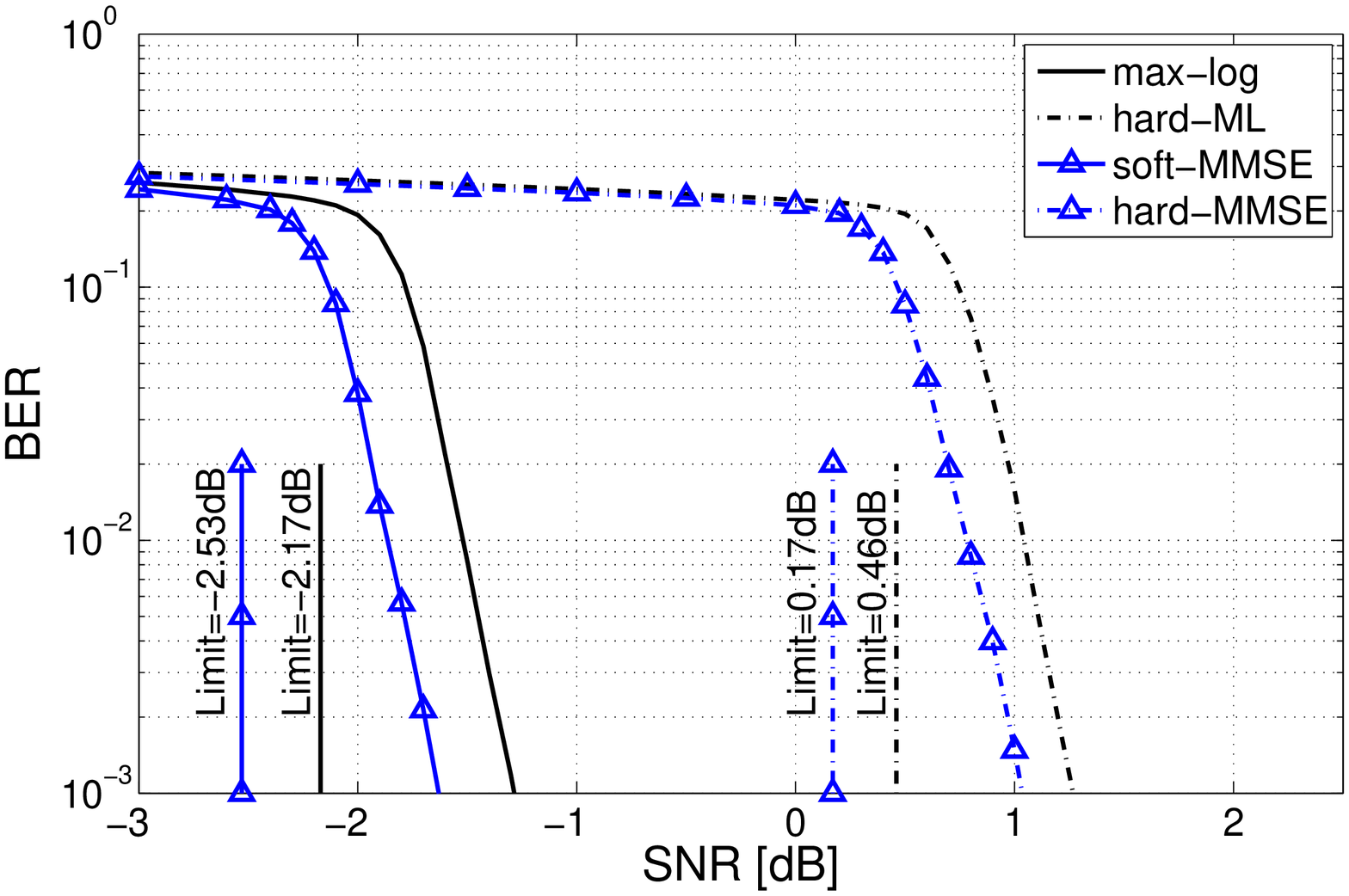}}
\hfill
\subfigure[]{\includegraphics[scale=0.363]{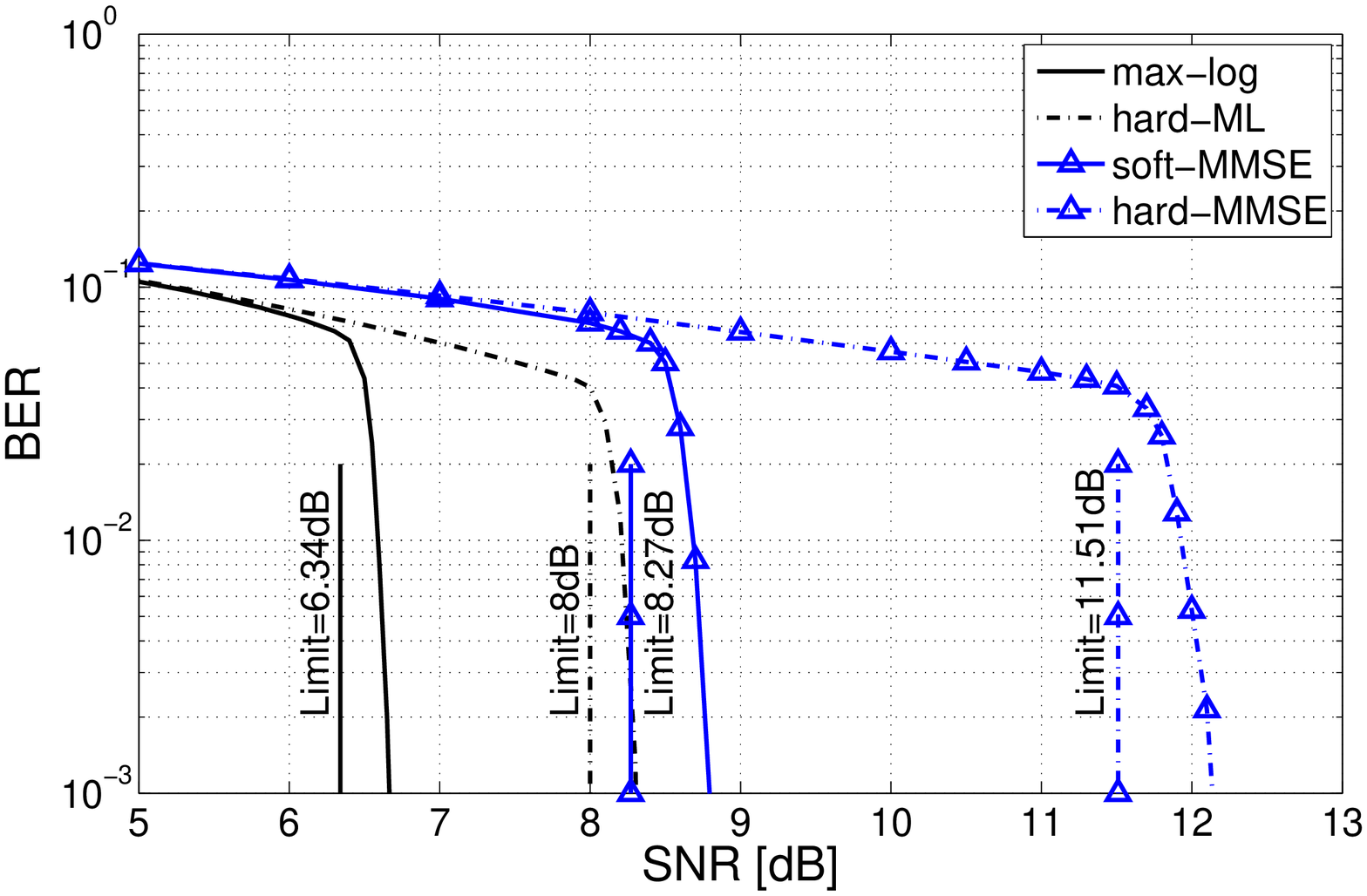}}
\vspace*{-9mm}
\caption{BER vs.\ SNR for a $4\!\times\!4$ MIMO system with Gray-labeled $4$-QAM and LDPC codes of (a) rate $1/4$ and (b) rate $3/4$.}
\label{fig:ber_basic_4x4_4QAM_gray}
\vspace*{-7mm}
\end{figure*}

The BERs obtained for code rates of $1/4$ ($2$\,bpcu) and $3/4$ ($6$\,bpcu) are shown in
Fig.~\ref{fig:ber_basic_4x4_4QAM_gray}(a) and (b), respectively.
Vertical lines indicate the respective capacity limits, i.e., the minimum SNR required for the target rate according to Fig.~\ref{fig:basic_gray}(a). 
It is seen that the LDPC code designs are less than $1\,$dB away from the capacity limits.
At low rates soft MMSE performs best and hard ML performs worst whereas at high rates max-log and hard MMSE  give the best and worst results, respectively.
More specifically, at rate ${1}/{4}$ soft MMSE outperforms max-log and hard ML demodulation by $0.3$\,dB and $2.9$\,dB, respectively (cf.~Fig.~\ref{fig:ber_basic_4x4_4QAM_gray}(a));
at rate ${3}/{4}$ soft MMSE performs $0.5$\,dB poorer than hard ML and $2.1\,$dB poorer than max-log (cf.~Fig.~\ref{fig:ber_basic_4x4_4QAM_gray}(b)).
These BER results confirm the capacity-based observation 
that there is no universal (i.e., rate- and SNR-independent) demodulator performance ranking.
We note that the block error rate results in \cite{Michalke:2006aa} imply similar conclusions, even though not explicitly mentioned in that paper.

\section{Other Demodulators}
\label{sec:other}
In the following, we study the system capacity of several other MIMO-BICM demodulators that differ in their underlying principle and their computational complexity.
Unless stated otherwise, capacity results shown in this section pertain to a $4\times4$ MIMO system with $4$-QAM using Gray labeling ($\slambda=8$).
The results for asymmetric $2\times4$ MIMO systems with $16$-QAM
(shown in \cite{Fertl:2009ac} but not here)
essentially confirm the general distinction between hard and soft demodulators
observed in connection with
Fig.~\ref{fig:basic_gray}(b). 

\subsection{List-based Demodulators}
In order to save computational complexity, \eqref{eq:max-log} can be approximated by decreasing the size of the search set, i.e., replacing $\shA$ with a smaller set.
Usually, this is achieved by generating a (non-empty) candidate list $\sL \subseteq \shA$
and restricting the search in \eqref{eq:max-log} to this list, i.e., 
\begin{equation}
\shLLRj = 
\frac{1}{\ssigw}\!\bigg[ {\underset{\sx\in\sL\cap\shAz}{\min} \|\sr\!-\!\sH\sx\|^2}\,-{\underset{\sx\in\sL\cap\shAo}{\min} \|\sr\!-\!\sH\sx\|^2}\bigg]. \label{eq:list}
\end{equation}
\NEW{As the number of operations required to compute the metric for each candidate of the list is $\sO(\sMt\sMr)$, the overall computational complexity of the metric evaluations and minima searches
in \eqref{eq:list} 
scales as \sloppy $\sO(\sMt\sMr \sLcand)$.} 
Thus, the list size $\sLcand$ allows to trade off performance for complexity savings.
A larger list size generally incurs higher complexity but yields more accurate approximations of the max-log LLRs.
For a fixed list size, the performance 
further depends on how the list $\sL$ is generated.
In the following, we consider two types of list generation, one based on
sphere decoding and the other on bit flipping.

\vspace{1mm}

\subsubsection{List Sphere Decoder (LSD)}
The LSD proposed in \cite{hochbrink03} uses a simple modification of the hard-decision sphere decoder \cite{damen03} to generate the candidate list $\sL$ 
such that it contains the $\sLcand$ symbol vectors $\sx$ with the smallest ML metric $\|\sr\!-\!\sH\sx\|^2$ (thus, by definition $\sL$ contains the hard ML solution $\shxml$ in \eqref{eq:hardML}).
If the $\sj$th bit in the labels of {\em all} $\sx\in\sL$ equals $1$, the set $\sL\cap\shAz$ is empty and
\eqref{eq:list} cannot be evaluated. Since in this case there is strong evidence for $\scj=1$ 
(at least if $\sLcand$ is not too small), the 
LLR $\shLLRj$
is set to a prescribed positive value $\sLLRLSD\gg 0$. Analogously, $\shLLRj=-\sLLRLSD$ in case $\sL\cap\shAo$ 
is empty.
While the LSD may offer significant complexity savings compared to max-log demodulation, statements about its computational complexity are difficult and depend strongly on the actual implementation of the sphere decoder as well as the choice of the list size \NEW{(for details we refer to \cite{hochbrink03})}. 
We note that the case $\sLcand\!=\!2^{\slambda}=|\sA^{\sMt}|$ implies $\sL=\sA^{\sMt}$; thus,
$\sL\cap\shAcj=\shAcj$ such that \eqref{eq:list}
equals the max-log demodulator in \eqref{eq:max-log}.
The other extreme is a list size of one, i.e., $\sL=\{\shxml\}$ (cf.\ \eqref{eq:hardML}), in which case
either $\sL\cap\shAz$ or $\sL\cap\shAo$ is empty (depending on the bit label of $\shxml$); here,
$\shLLRj=(2\shcj-1)\sLLRLSD$ where
$\svhc=(\shcbit_1\dots \shcbit_{\slambda})\strans=\smu^{-1}(\shxml)$
and thus the LSD output is equivalent to hard ML demodulation (except for the choice of
$\sLLRLSD$, which is irrelevant, however, for capacity).

\vspace{1mm}

\begin{figure}
\centering
\includegraphics[scale=0.365]{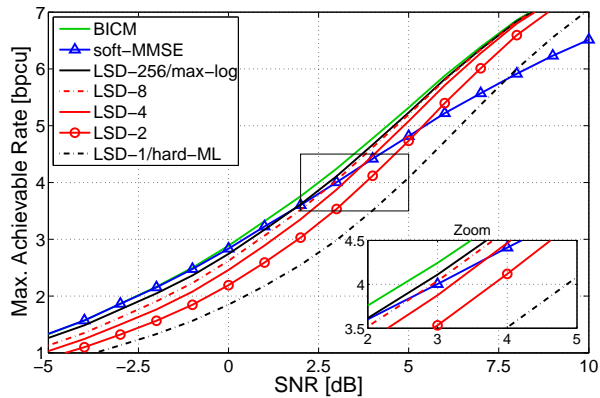}
\vspace*{-6mm}
\caption{System capacity of LSD with list size $\sLcand\in\{1,2,4,8,256\}$ ($4\!\times\!4$ MIMO, $4$-QAM, Gray labeling).}
\label{fig:LSD_4x4_4QAM_gray}
\vspace*{-6mm}
\end{figure}

{\em Capacity Results.}
Fig.~\ref{fig:LSD_4x4_4QAM_gray} shows the maximum rates achievable with an LSD for various list sizes.
BICM and soft MMSE capacity are shown for comparison.
Note that with $4$-QAM and $\sMt=4$, $\sLcand=256$ and $\sLcand=1$ correspond to max-log and hard ML demodulation,
respectively. It is seen that with increasing list size the gap between LSD and max-log decreases rapidly,
specifically at high rates.
In particular, the LSD with list sizes of $\sLcand\!\ge\!8$ is already quite close to max-log performance. However, at low rates LSD (even with large list sizes) is outperformed by soft MMSE:
below $5.3$\,dB, $3.7$\,dB, and $2.8$\,dB the system capacity of soft MMSE is higher than that of LSD with list size $2$, $4$, and $8$, respectively.
Similar observations apply to other antenna configurations and symbol constellations (see \cite{Fertl:2009ac}).

\vspace{1mm}

\subsubsection{Bit Flipping Demodulators}\label{sssec:flipping}
Another way of generating the candidate list $\sL$, proposed in \cite{Wang:2006aa}, is to flip some of the bits in the label of the hard ML symbol vector estimate $\shxml$ in \eqref{eq:hardML}. More generally, the
ML solution $\shxml$ can be replaced by a symbol vector $\shx\in\shA$ obtained with an arbitrary
hard-output demodulator (e.g., hard ZF and MMSE demodulation).
Let $\svhc = 
\smu^{-1}(\shx)$
denote the 
bit label of $\shx$.
The candidate list then consists of all symbol vectors whose bit label has Hamming distance
at most $\sDham\le\slambda$ from $\svhc$, 
i.e.,
$\sL =  \{\sx\!:\,\shamming(\smu^{-1}(\sx),\svhc) \le \sDham\}$. Here,
$\shamming(\svc_1,\svc_2)$ denotes the Hamming distance between two bit labels $\svc_1$ and $\svc_2$.
This list can be generated by systematically flipping up to $\sDham$ bits in $\svhc$ and mapping the results to symbol vectors.
The resulting list size is given by $\sLcand\!=\!\sum_{\sdi=0}^{\sDham}\binom{\slambda}{\sdi}$. Here, the structure of the list generated with bit flipping allows to reduce the complexity per candidate to $\sO(\sMr)$,\NEW{\footnote{\NEW{Changing the value of a particular bit changes only one symbol in the symbol vector. Thus, the residual $\sr-\sH\sx$ in \eqref{eq:list} can be easily updated by adding an appropriately scaled column of $\sH$. This requires only $\sO(\sMr)$ instead of $\sO(\sMt\sMr)$ operations.}}}  giving an overall complexity of $\sO(\sMr\sLcand)$ (plus the operations required for the initial estimate). 
For $\sDham=\slambda$, $\sL=\sA^{\sMt}$ and \eqref{eq:list} reduces to max-log demodulation;
furthermore, with $\shx=\shxml$ and $\sDham=0$ there is $\sL=\{\shxml\}$ and \eqref{eq:list}
becomes equivalent to hard ML demodulation.
In contrast to the LSD, bit flipping with $\sDham>0$ ensures that $\sL\cap\shAz$
and $\sL\cap\shAo$ are nonempty
so that \eqref{eq:list} can always be evaluated. 

\vspace{1mm}

\begin{figure*}[t]
\centering
\subfigure[]{\includegraphics[scale=0.365]{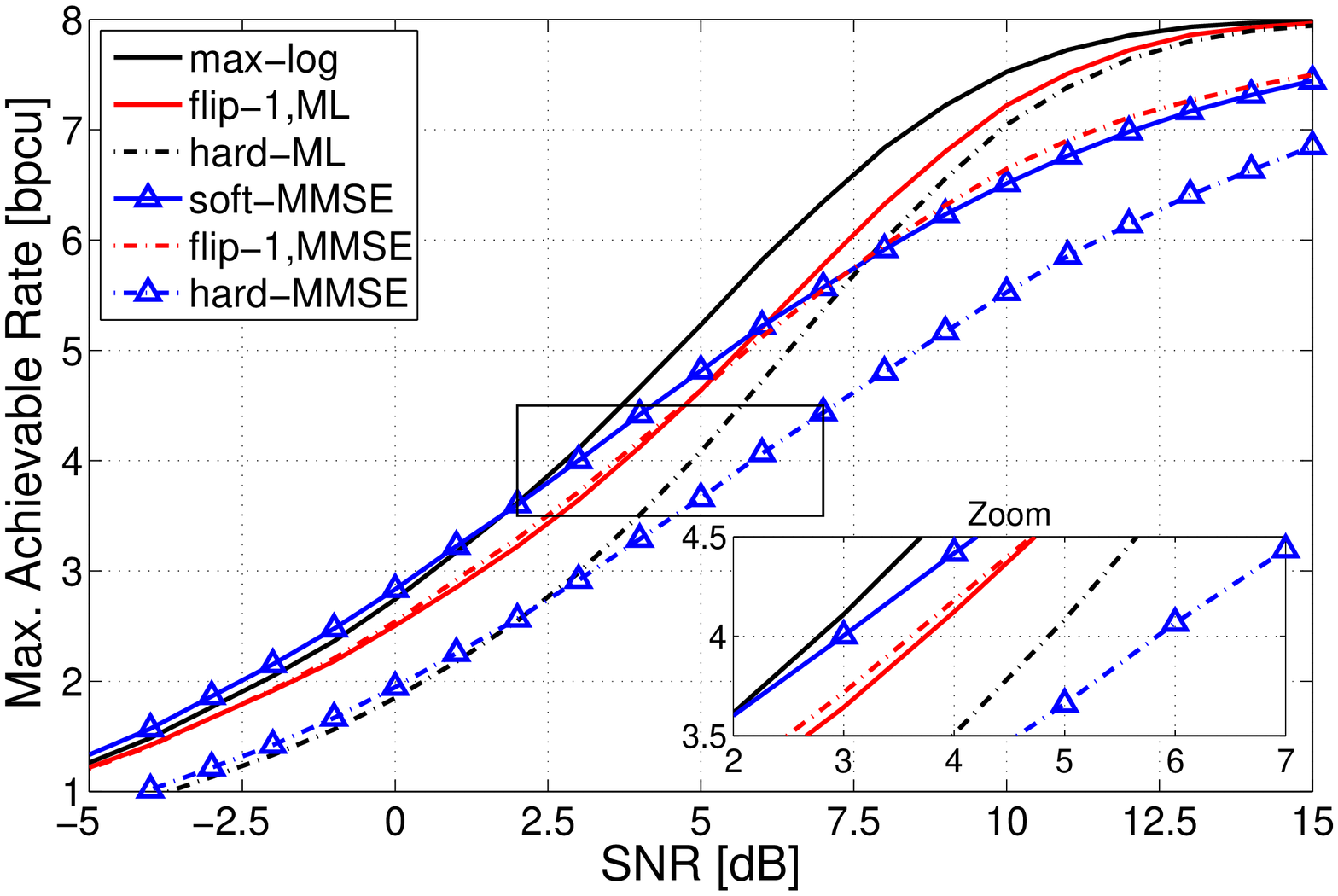}}
\hfill
\subfigure[]{\includegraphics[scale=0.365]{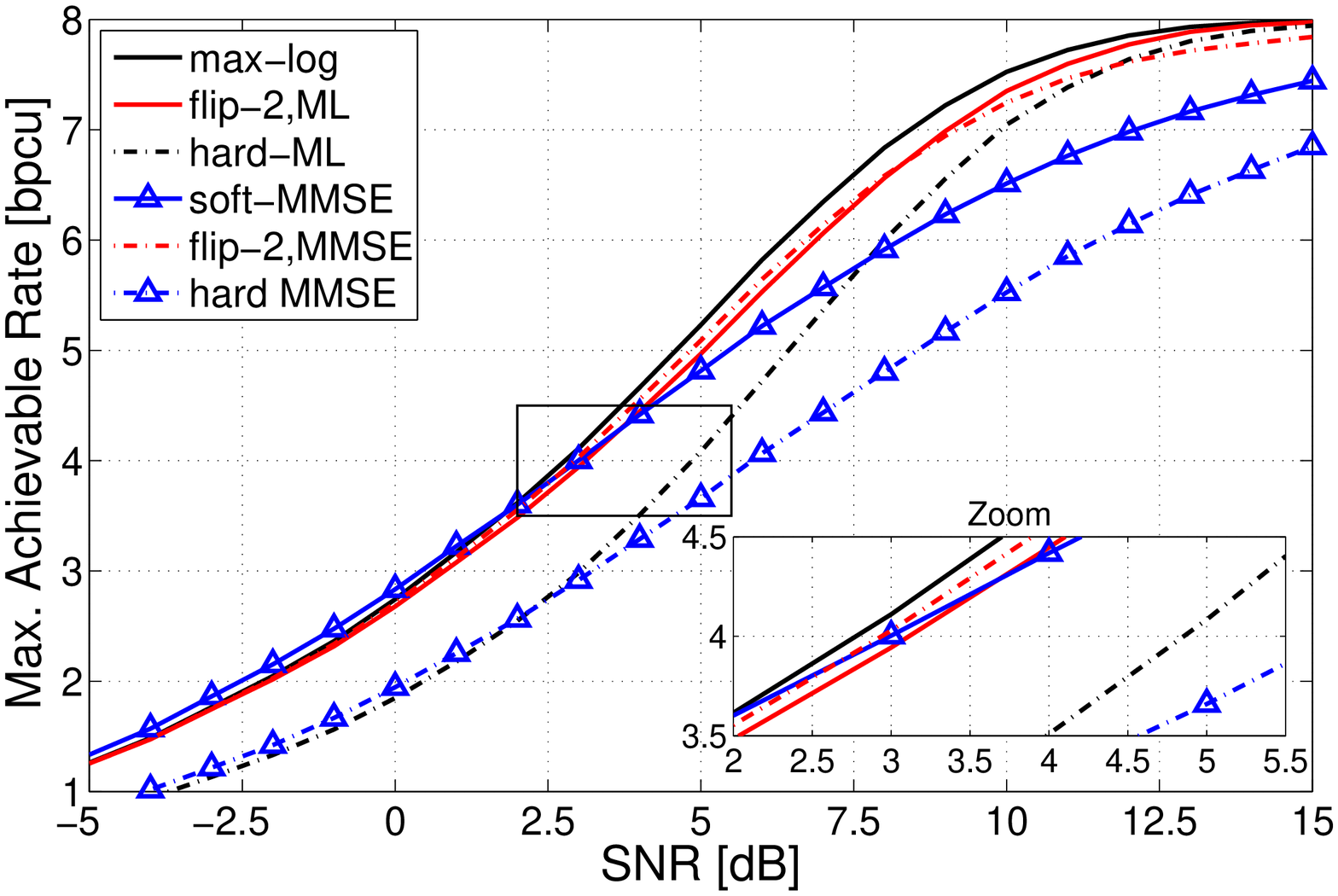}}
\vspace*{-3mm}
\caption{\label{fig:flip_4x4_4QAM_gray}System capacity of bit flipping demodulator with (a) $\sDham=1$ and (b) $\sDham=2$
($4\!\times\!4$ MIMO, $4$-QAM, Gray labeling).}
\vspace*{-7mm}
\end{figure*}

{\em Capacity Results.}
Fig.~\ref{fig:flip_4x4_4QAM_gray} shows the maximum rates achievable with bit flipping demodulation where the initial symbol vector estimate is chosen either as the hard ML solution $\shxml$ in \eqref{eq:hardML} or the hard MMSE estimate $\sQuant(\symmse)$ (cf.\ \eqref{eq:wiener}).
For $\sDham=1$ ($\sLcand=9$), Fig.~\ref{fig:flip_4x4_4QAM_gray}(a) reveals that flipping 1 bit (labeled 'flip-1') allows for significant performance improvements over the respective initial hard demodulator
(about $2.1$\,dB at $2$\,bpcu).
For rates below $5$\,bpcu, hard ML and hard MMSE initialization yield effectively identical results, 
with a maximum loss of $0.9$\,dB (at $3.5$\,bpcu) compared to soft MMSE.
At higher rates, MMSE-based bit flipping outperforms soft MMSE demodulation slightly. 
For $\sDham=2$ ($\sLcand=37$), it can be seen from Fig.~\ref{fig:flip_4x4_4QAM_gray}(b) that bit flipping demodulation performs close to max-log below $4$\,bpcu and that hard ML and hard MMSE initialization are very close to each other for almost all rates and SNRs
; in fact, below $6.7$\,bpcu hard MMSE initialization performs slightly better than hard ML initialization while at higher rates ML initialization gives slightly better results.
\NEW{To maintain this behavior for larger constellations and more antennas, the maximum Hamming distance $\sDham$ may have to increase with increasing $\slambda$ (see \cite{Fertl:2009ac}).}


\subsection{Lattice-Reduction-Aided Demodulation}
Lattice reduction (LR) is 
an important technique for improving the performance or complexity of MIMO demodulators \cite{yao02,wubb04b}
for the case of QAM constellations. The basic underlying idea is to view the columns of the channel matrix $\sH$ as basis vectors of a point lattice.
LR then yields an alternative basis which amounts to a
transformation of the system model \eqref{eq:system-model} prior to demodulation;
the advantage of such an approach is that the
transformed channel matrix (i.e., the reduced basis) has improved properties (e.g., smaller condition number).
An efficient algorithm to obtain a reduced basis 
was proposed by Lenstra, Lenstra, and Lov{\'a}sz (LLL) \cite{lenstra82}.
The overall computational complexity of LR-aided demodulation depends 
on the complexity of the LLL algorithm which \NEW{is currently an active research topic. Bounds on the average computational complexity of the LLL algorithm have been provided in \cite{Jalden:2008aa}.}
A comparison of different LR methods in the context of MIMO hard demodulation was provided in \cite{Wubben:2007aa}.

Since LR algorithms are often formulated for equivalent real-valued models, 
we assume for now that all quantities are real-valued.
Any lattice basis transformation is described by 
a unimodular transformation matrix $\sT$, i.e., a matrix with integer entries and $\det(\sT)=\pm 1$.
Denoting the ``reduced channel'' by $\sHtilde=\sH\sT$ and defining $\sz=\sT^{-1}\sx$, the system model  \eqref{eq:system-model} can be rewritten as
\begin{equation}
\sr = \sH\sx +\sw = \sHtilde\sz+\sw.
\label{eq:red_system}
\end{equation}
Under the assumption $\sx \in \sZint^{\sMt}$ (which for QAM can be ensured by an appropriate offset and scaling), the unimodularity of $\sT$ guarantees
$\sz \in \sZint^{\sMt}$ and hence any demodulator can be applied to the better-behaved transformed system model on the right-hand side of \eqref{eq:red_system}.
LR-aided soft demodulators (cf.~\cite{Windpassinger:2003aa}) are essentially list-based
\cite{Silvola:2006aa,Ponnampalam:2007aa}, 
and often apply bit flipping (cf.~Section \ref{sssec:flipping}) to LR-aided hard-output demodulators.
Here, we restrict to LR-aided hard and soft output MMSE demodulation \cite{wubb04b}.

\vspace{1mm}

\begin{figure}
\centering
\includegraphics[scale=0.365]{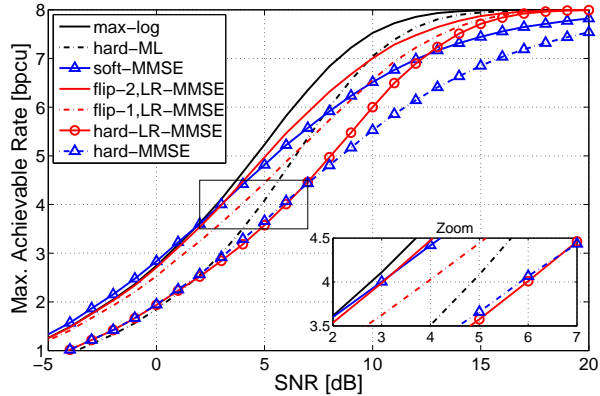}
\vspace*{-6mm}
\caption{System capacity of LR-aided hard and soft MMSE demodulation ($4\!\times\!4$ MIMO, $4$-QAM, Gray labeling).}
\vspace*{-7mm}
\label{fig:LR_4x4_4QAM_gray}
\end{figure}

{\em Capacity Results.}
Fig.~\ref{fig:LR_4x4_4QAM_gray} shows the capacity results for hard and soft LR-aided MMSE demodulation.
Soft outputs are obtained by applying bit flipping with $\sDham=1$ and $\sDham=2$ to the LR-aided hard MMSE demodulator output (cf.~Section \ref{sssec:flipping}).
It is seen for $4\times4$ MIMO with $4$-QAM ($\slambda=8$) in Fig.~\ref{fig:LR_4x4_4QAM_gray} that LR with hard MMSE demodulation shows a significant performance advantage over hard MMSE demodulation for SNRs above $7.2$\,dB (rates higher than $4.5$\,bpcu).
At rates higher than about $7.1$\,bpcu, LR-aided hard demodulation even outperforms soft MMSE demodulation.
Bit flipping is helpful particularly at low-to-medium rates. Thus, for SNRs below $6.8$\,dB (rates lower than $5.2$\,bpcu) LR-aided soft demodulation with $\sDham=1$ essentially performs better than hard ML.
When flipping up to $\sDham=2$ bits, LR-aided soft demodulation closely approaches max-log performance and reveals a significant performance advantage over soft MMSE demodulation without LR in the high-rate regime.

\subsection{Semidefinite Relaxation Demodulation}
Based on convex optimization techniques, semidefinite relaxation (SDR) is an approach to approximately solve the hard ML problem \eqref{eq:hardML} with polynomial worst-case complexity \cite{Tan:2001aa,wong_sp02}. 
We specifically consider hard-output and soft-output versions of an SDR demodulator that approximates max-log demodulation and has an overall worst-case complexity of $\sO(\slambda^{4.5})$ (see \cite{Steingrimsson:2003aa}). 
We note that this approach applies only to BPSK or 4-QAM alphabets
and employs a randomization procedure described in detail in \cite{wong_sp02}.

\begin{figure}
\centering
\includegraphics[scale=0.365]{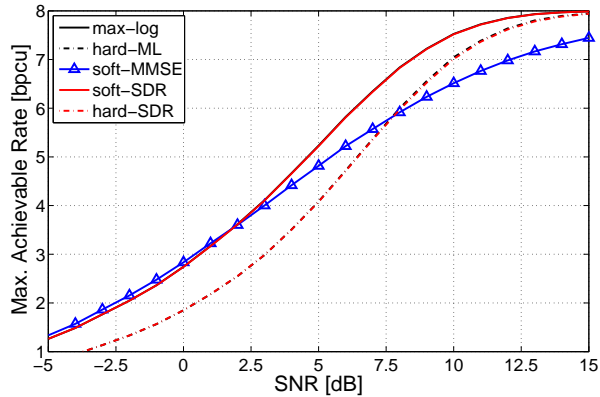}
\vspace*{-6mm}
\caption{System capacity of hard and soft SDR demodulation ($4\!\times\!4$ MIMO, $4$-QAM, Gray labeling).
The curves for hard and soft SDR are identical to those for hard ML and max-log, respectively.}
\vspace*{-7mm}
\label{fig:SDR_4x4_4QAM_gray}
\end{figure}

{\em Capacity Results.}
In Fig.~\ref{fig:SDR_4x4_4QAM_gray} we show the system capacity 
for a $4\!\times\!4$ MIMO system with $4$-QAM ($\slambda=8$)
using hard and soft SDR demodulation (as described in \cite{Steingrimsson:2003aa}) and randomization with $25$ trials.
Surprisingly, 
hard and soft SDR demodulation here exactly match the performance of hard ML and max-log demodulation, respectively. 

\subsection{Infinity-Norm Demodulator}
The VLSI implementation complexity of the sphere decoder for hard ML demodulation 
is significantly reduced by replacing the $\sltwo$ norm in \eqref{eq:hardML} with the $\slinf$ norm, i.e.,
\begin{equation}
\shxlinf = \underset{\sx\in\shA}{\arg \; \min}\, \|\sQqr\sherm\,\sr-\sRqr\sx\|_\infty . \label{eq:hardlinf}
\end{equation}
Here, the $\ell^\infty$ norm is defined as
$\|\mathbf{a}\|_\infty\define\text{max}\big\{\text{Re}\{a_1\},\dots,\text{Re}\{a_M\},\text{Im}\{a_1\},\dots,\text{Im}\{a_M\}\big\}$
and $\sQqr$ and $\sRqr$
are the $\sMr\!\times\!\sMt$ unitary and $\sMt\!\times\!\sMt$ upper triangular factors
in the QR decomposition $\sH = \sQqr \sRqr$ of the channel matrix.
The advantage of \eqref{eq:hardlinf} is that 
expensive squaring operations are avoided \NEW{and fewer nodes are visited during the tree search 
underlying the sphere decoder \cite{Burg:2005aa,Seethaler:2010aa}.}
If \eqref{eq:hardlinf} has no unique solution, 
one solution is selected at random.

Soft outputs can be generated by using the $\slinf$-norm sphere decoder to determine
\begin{equation*}
\tilde\sx_\sj^b = \underset{\sx\in\shAcj}{\arg \; \min}\, \|\sQqr\sherm\,\sr-\sRqr\sx\|_\infty  
\end{equation*}
for $b \in \{0,1\}$ and then evaluating the approximate LLRs
using the $\sltwo$ norm:
\begin{equation*}
\shLLRj = 
\frac{1}{\ssigw}\!\Big[  \|\sr-\sH\tilde\sx_\sj^0\|^2\,-\|\sr-\sH\tilde\sx_\sj^1\|^2\Big]. 
\end{equation*}

\begin{figure*}
\centering
\subfigure[]{\includegraphics[scale=0.365]{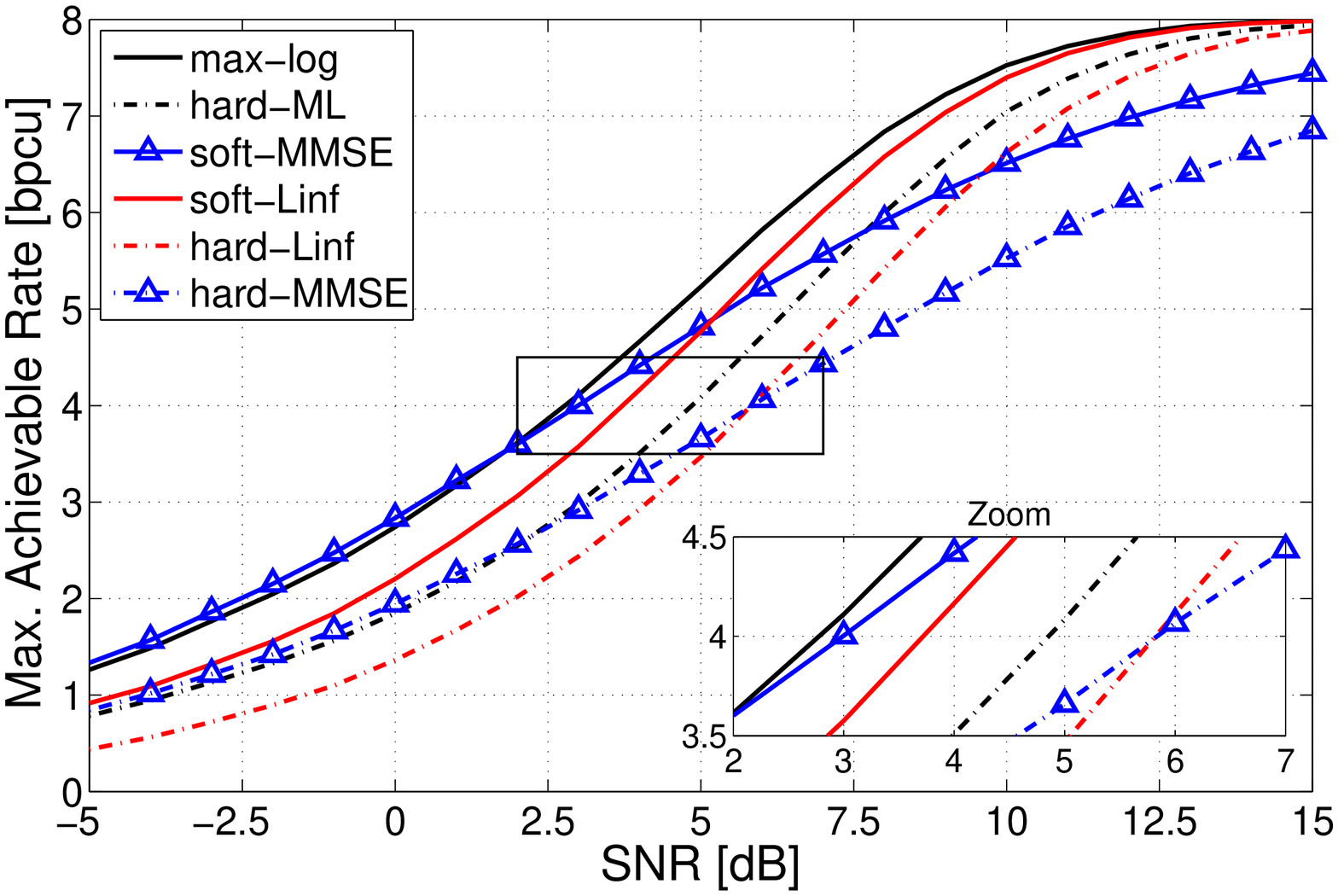}}
~\hfill~
\subfigure[]{\includegraphics[scale=0.365]{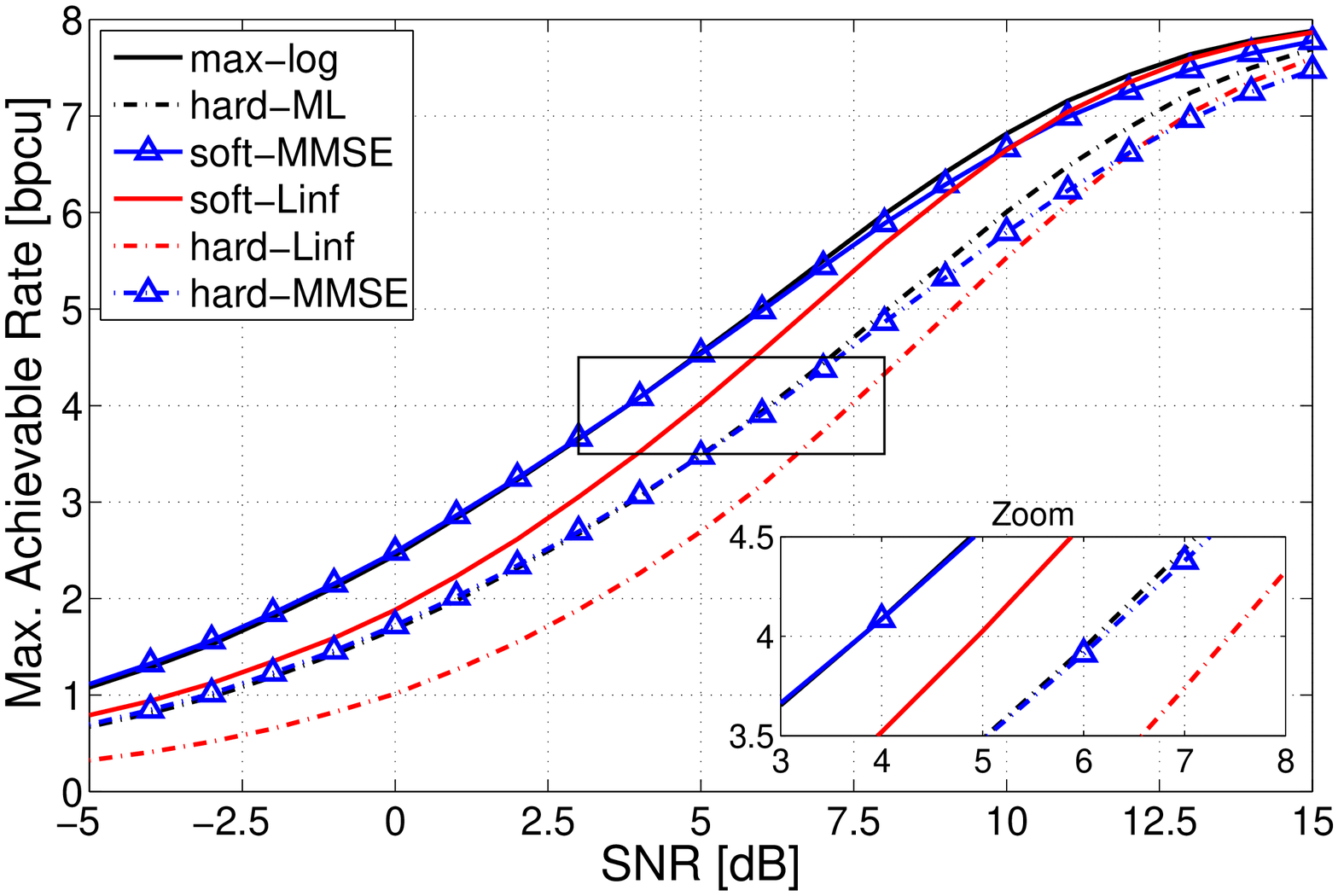}}
\vspace*{-10mm}
\caption{\label{fig:Linf_4x4_4QAM_gray}System capacity of hard and soft $\slinf$-norm demodulation for
(a) $4\!\times\!4$ MIMO with $4$-QAM and
(b) $2\!\times\!4$ MIMO with $16$-QAM (Gray labeling in both cases).}
\vspace*{-7mm}
\end{figure*}

{\em Capacity Results.}
Fig.~\ref{fig:Linf_4x4_4QAM_gray} shows the system capacity for hard and soft $\slinf$-norm demodulation.
For the $4\times 4$ case with $4$-QAM in Fig.~\ref{fig:Linf_4x4_4QAM_gray}(a),
hard and soft $\slinf$-norm demodulation perform within 1\,dB of hard ML and max-log, respectively. 
At rates below about $4$\,bpcu, $\slinf$-norm demodulation is outperformed by 
MMSE demodulation, though. 
For the $2\times 4$ case with 16-QAM depicted in Fig.~\ref{fig:Linf_4x4_4QAM_gray}(b), all soft-output baseline demodulators perform almost identical and the same is true for all hard-output baseline demodulators, i.e., there is only a distinction between soft and hard demodulation (cf.~Fig.~\ref{fig:basic_gray}(b)).
However, soft and hard $\slinf$-norm demodulation perform significantly worse in this asymmetric setup, specifically at low-to-medium rates. At $2$\,bpcu, soft $\slinf$-norm demodulation requires $1.75$\,dB higher SNR than max-log and soft MMSE and hard $\slinf$-norm demodulation requires $2.3$\,dB higher SNR than hard ML/MMSE.

\vspace*{-2mm}
\subsection{Successive and Soft Interference Cancelation}

Successive interference cancelation (SIC) is a hard-output demodulation approach that became popular with
the V-BLAST ({\em Vertical Bell Labs Layered Space-Time}) system 
\cite{wolniansky98}.
Within one SIC iteration, only the layer with the largest post-equalization SNR
is detected and its contribution to the receive signal is subtracted (canceled).
A SIC implementation
that replaces the ZF detector from \cite{wolniansky98} with an MMSE detector and orders the layers efficiently according to signal-to-interference-plus-noise-ratio (SINR) was presented in \cite{hassibi00}.
\NEW{Note that this approach shows a complexity order of $\sO(\sMr\sMt^2)$ which is the same as for linear MIMO demodulation.}
Suboptimal but more efficient SIC schemes
are discussed in \cite{wubb03}.


\def\siter{j}

A parallel soft interference cancelation (SoftIC) scheme with reduced
error propagation 
was proposed in \cite{choi_wcnc00}. 
SoftIC is an iterative method that
iteratively performs
(i) parallel MIMO interference cancelation based on soft symbols and
(ii) computation of improved soft symbols using the output of the interference cancelation stage.
The complexity of one SoftIC iteration 
depends linearly on the number of antennas.
Here, we use a modification that builds upon bit-LLRs. 
Let $\shLLRik[\siter]$ denote the LLR for the $\si$th bit in layer $\sk$ obtained in the $\siter$th iteration.
Symbol probabilities can then be obtained as
\[
\sPkj(\sxi) = \prod_{\si=1}^{\sm}
\frac{ \exp\left(b_\si(\sxi)\shLLRik[\siter]\right) }{ 1+\exp\left(\shLLRik[\siter]\right) },
%
\]
with $b_\si(\sxi)$ denoting the $\si$th bit in the label of $\sxi\in\sA$, leading to the
soft symbol estimate
\[
\sxsoft_\sk^{(\siter)}    = \sum_{\sxi\in\sA} \sxi\,\sPkj(\sxi)\,. 
\]
Soft interference cancelation for each layer then yields
\begin{equation}\label{eq:intcancel}
\sr_\sk^{(\siter)} = \sr - \sum_{\sk'\ne \sk} \sh_{\sk'}\sxsoft_{\sk'}^{(\siter)}
= \sh_\sk\, \sxi_\sk + \sum_{\sk'\ne \sk} \sh_{\sk'} \bigl(\sxi_{\sk'} - \sxsoft_{\sk'}^{(\siter)}\bigr) + \sw,
\end{equation}
where $\sh_\sk$ denotes the $\sk$th column of $\sH$.
Finally, updated LLRs $\shLLR_\sk^{(\si)}[\siter\!+\!1]$ are calculated from \eqref{eq:intcancel}
based on a Gaussian assumption for the residual interference plus noise (for details we refer to \cite{choi_wcnc00}).
%
In contrast to \cite{choi_wcnc00}, we suggest to initialize the scheme with the LLRs obtained by a low-complexity soft demodulator, e.g., the soft ZF demodulator in Section \ref{ssec:lin_demod}. 
\NEW{By carefully counting operations it can be shown that}
the complexity per iteration of the above SoftIC algorithm scales as $\sO(2^\sm\sMt(\sm+\sMr))$.

\vspace{1mm}

\begin{figure}
\centering
\subfigure[]{\includegraphics[scale=0.365]{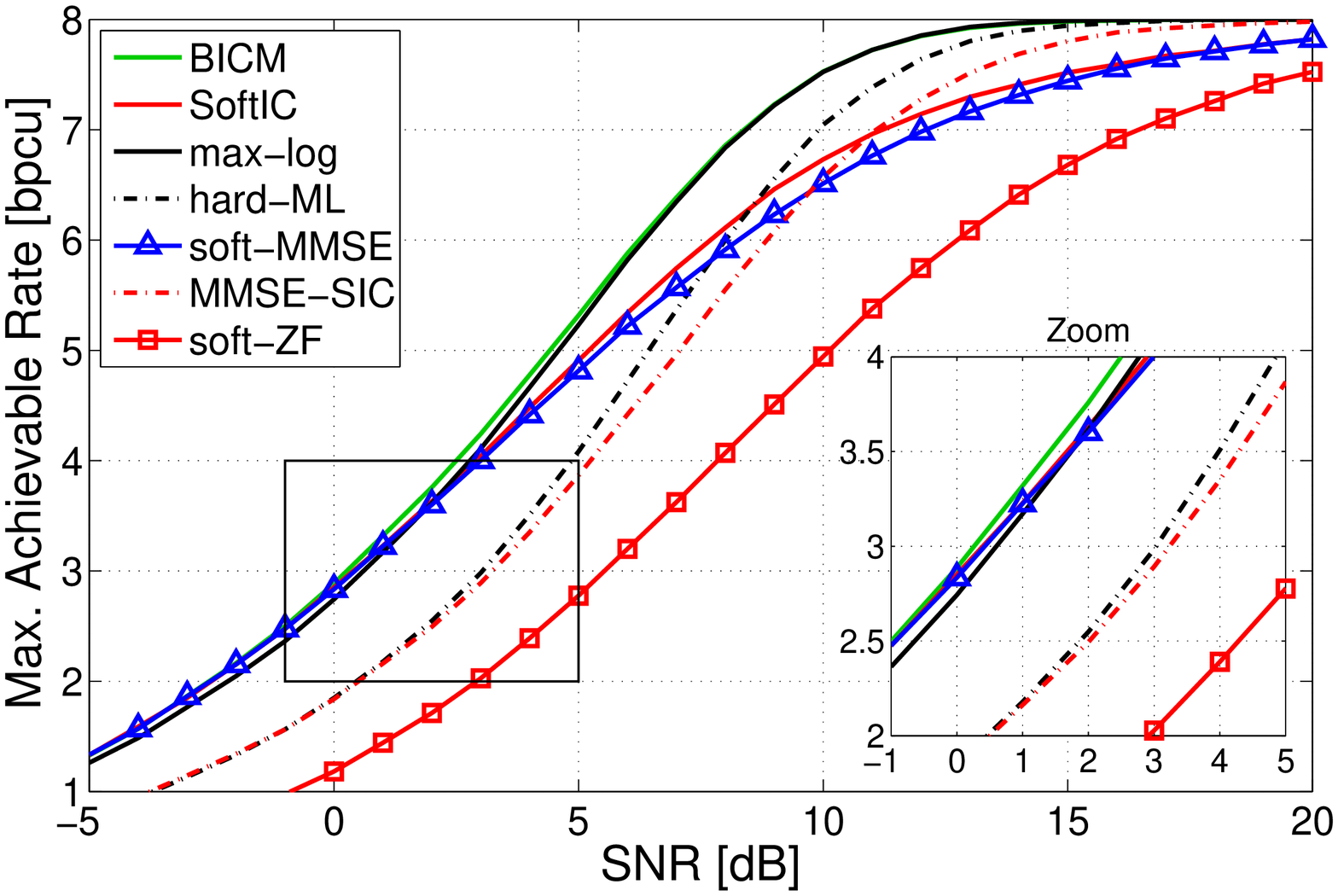}}
~\hfill~
\subfigure[]{\includegraphics[scale=0.365]{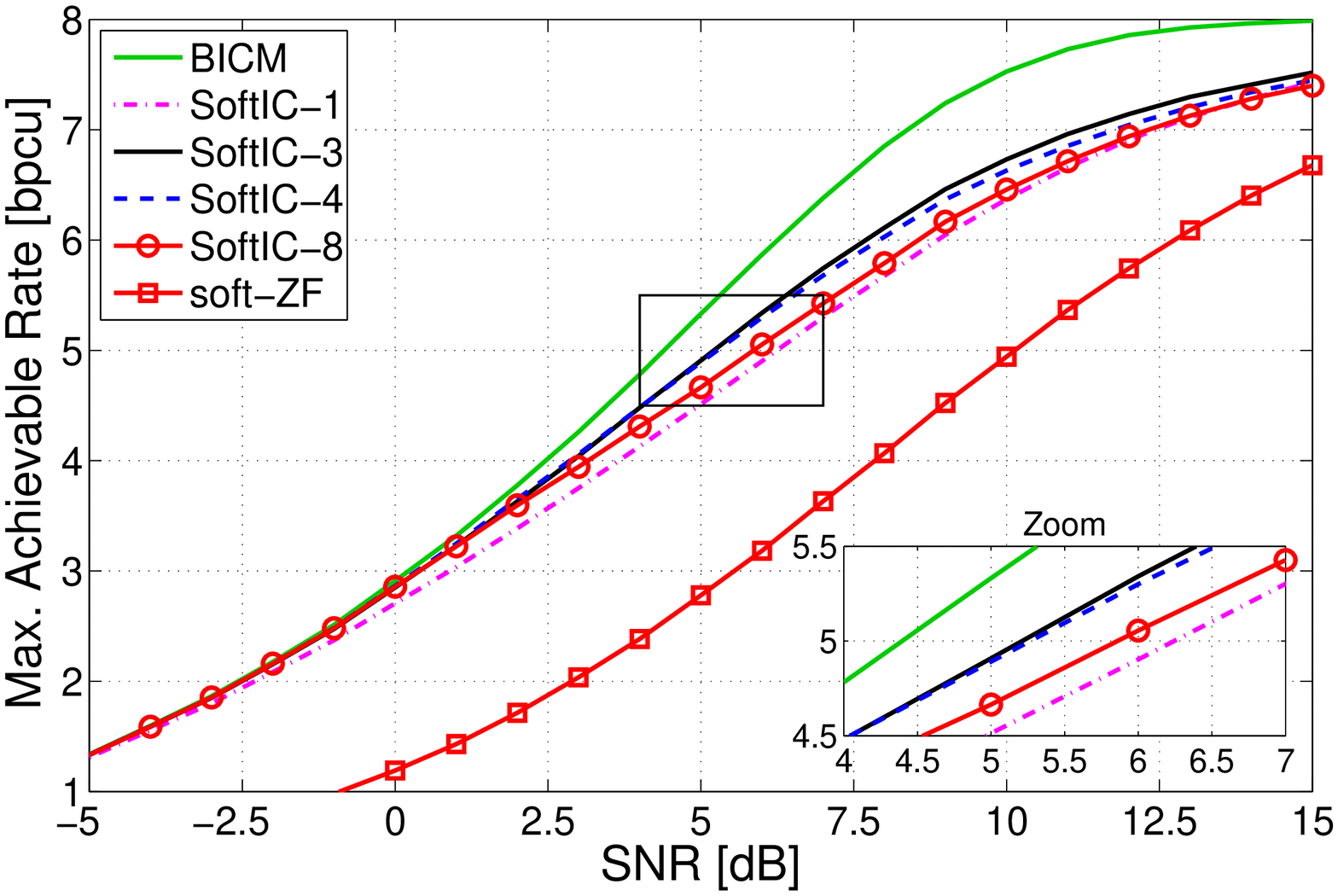}}
\vspace*{-9mm}
\caption{System capacity of MMSE-SIC and SoftIC ($4\!\times\!4$ MIMO, $4$-QAM, Gray labeling).}
\label{fig:VBLAST_SIC_4x4_4QAM_gray}
\vspace*{-7mm}
\end{figure}

{\em Capacity Results.}
In Fig.~\ref{fig:VBLAST_SIC_4x4_4QAM_gray}(a), we display capacity results for (hard) MMSE-SIC with detection ordering as in \cite{wubb03} (therein referred to as `MMSE-BLAST') and for SoftIC demodulation \NEW{with $3$ iterations} (initialized using a soft ZF demodulator whose performance is shown for reference).
Hard MMSE-SIC demodulation is seen to perform similarly to hard ML demodulation at low rates and even outperforms it slightly at very low rates.
At high rates, MMSE-SIC shows a noticeable gap to hard ML but outperforms soft MMSE and SoftIC. 
SoftIC is superior to MMSE-SIC for rates of up to $7$\,bpcu (SNRs below $11\,$dB).
At low rates, SoftIC even performs slightly better than max-log demodulation and
essentially coincides with BICM capacity and soft MMSE.
For the chosen system parameters, SoftIC closely matches soft MMSE at low rates and even outperforms it at high rates.
This statement is not generally valid, however.
For example, with $16$-QAM SoftIC performs 
poorer than soft MMSE 
even at high rates
(see \cite{Fertl:2009ac}).

\NEW{%
At high SNRs, we observed that SoftIC performance degrades if iterated too long (see Fig.~\ref{fig:VBLAST_SIC_4x4_4QAM_gray}(b), showing SoftIC with $1$, $3$, $4$, and $8$ iterations). This can be explained by the fact that at high SNRs 
the residual interference-plus-noise 
becomes very small and hence the LLR magnitudes grow unreasonably large. 
Our simulations showed that SoftIC performs best when terminated after $2$ or $3$ iterations.}



\section{Imperfect Channel State Information}
\label{sec:imperfect}
We next investigate the ergodic system capacity \eqref{eq:measure-exact} for the
case of imperfect channel state information (CSI).
In particular, we consider training-based estimation of the channel matrix $\sH$ and the noise variance $\ssigw$ and assess how the amount of training influences the performance of the various demodulators.

\vspace{1mm}

{\em Training-based Channel Estimation.}
To estimate the channel, the transmitter sends $\sNp>\sMt$ training vectors\footnote{While $\sNp=\sMt$ is sufficient to estimate $\sH$, extra training is required for estimation of $\ssigw$.}
which are arranged into an $\sMt\times\sNp$ training matrix $\sXp$.
We assume that $\sXp$ has full rank and has Frobenius norm \cite{Horn99} 
$\|\sXp\|^2\sfrob=\sNp\sPt$ such that
the power per channel use for training is the same as for the data.
Assuming that the channel stays constant for the duration of one block (which contains training and actual data),
the $\sMr\times\sNp$ receive matrix $\sYp$ induced by the training is given by
\begin{equation}
\sYp  = \sH \,\sXp  + \sV.
\label{eq:pilot-model}
\end{equation}
Here, $\sV$ is an $\sMr\times\sNp$ i.i.d.\ Gaussian noise matrix. 

Using \eqref{eq:pilot-model}, the least-squares (ML) channel estimate 
is computed as \cite{biguesh06}
\begin{equation}
\sHhat = \sYp\sXp\sherm(\sXp\sXp\sherm)^{-1}. \label{eq:Hest}
\end{equation}
This estimate is unbiased and its mean square error equals
\begin{equation*}
\sE\big\{ \| \sHhat - \sH \|^2\sfrob \big\} = \sMr\ssigw\, \text{tr}\big\{ (\sXp\sXp\sherm)^{-1} \big\} \ge \frac{\sMr\sMt^2}{\sNp}\frac{1}{\sSNR}\,,
\end{equation*}
where the lower bound is attained with orthogonal training sequences,
i.e., $\sXp\sXp\sherm=\frac{\sNp\sPt}{\sMt}\,\seye$ (we recall that $\sSNR=\sPt/\ssigw$ denotes the SNR).
The noise variance is then estimated as the mean power of $\sYp$ in the $(\sNp\!-\!\sMt)$-dimensional orthogonal complement of the range space of $\sXp\sherm$, i.e.,
\begin{equation}
\ssighat = \frac{1}{\sMr(\sNp\!-\!\sMt)}\|\sYp-\sHhat\sXp\|^2_{\text{F}}. \label{eq:sigest}
\end{equation}
The noise variance estimate is unbiased and its MSE is independent of the transmit power:
\begin{equation*}
\sE\big\{ | \ssighat - \ssigw |^2 \big\} = \frac{\sigma_\swi^4}{\sMr(\sNp\!-\!\sMt)}.
\end{equation*}

\vspace{1mm}

{\em Capacity Results.}
We show results for the ergodic system capacity of
mismatched\footnote{One could also modify these demodulators in order to take into account the fact that the CSI is imperfect as e.g.\ in \cite{Taricco:2005aa}; however, this is beyond the scope of this paper.} 
max-log, hard ML, and soft MMSE demodulation where the
true channel matrix and noise variance are replaced
by $\sHhat$ in \eqref{eq:Hest} and $\ssighat$ in \eqref{eq:sigest}, respectively.
Throughout, a $4\times4$ MIMO system with $4$-QAM and Gray labeling is considered ($\slambda\!=\!8$). Results for other demodulators with imperfect CSI are provided in \cite{Fertl:2009ac}.

\begin{figure*}
\psfrag{Number of Pilots}{\fontsize{7.5pt}{1}\sf Training Length}
\subfigure[]{\includegraphics[scale=0.355]{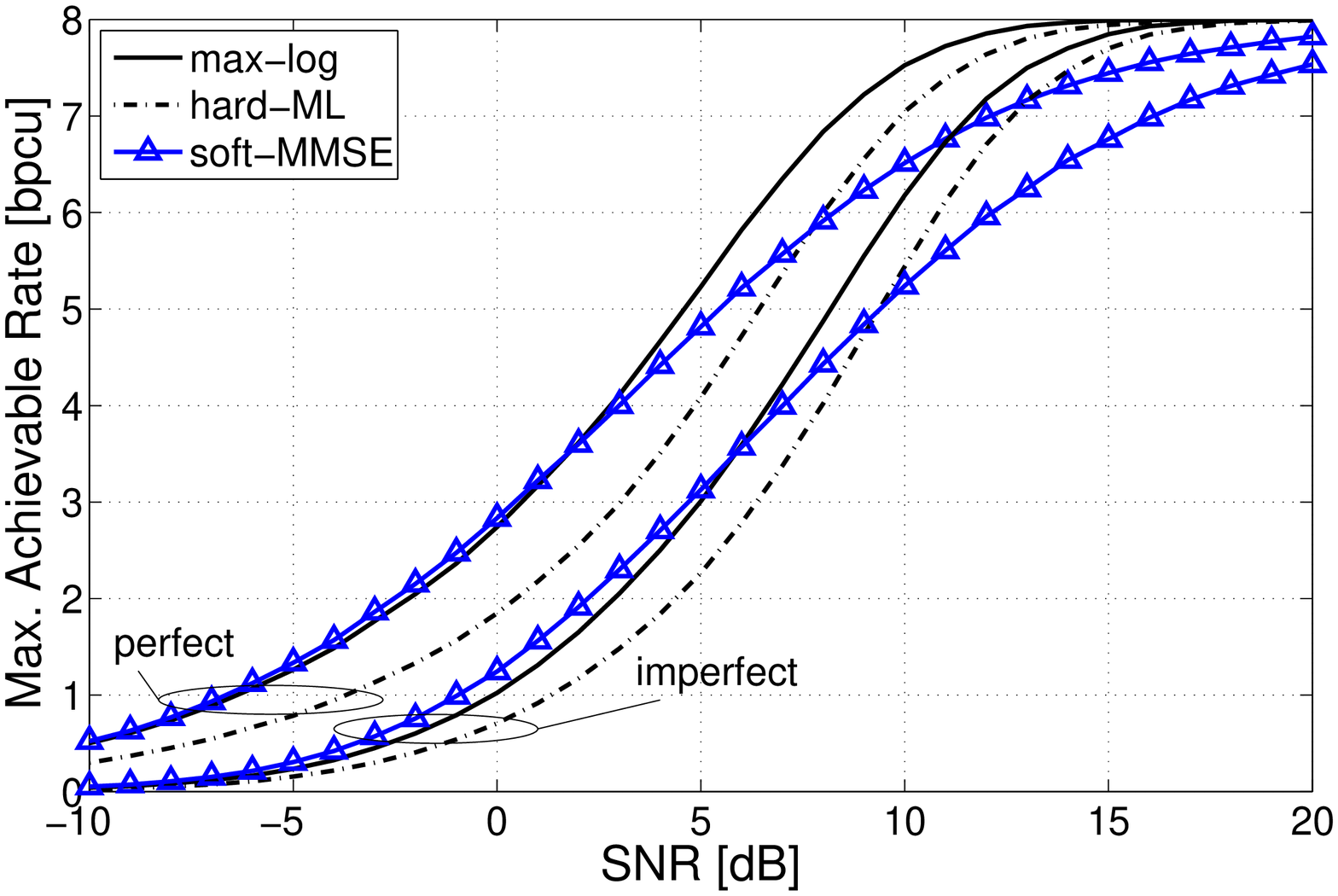}}
\subfigure[]{\includegraphics[scale=0.355]{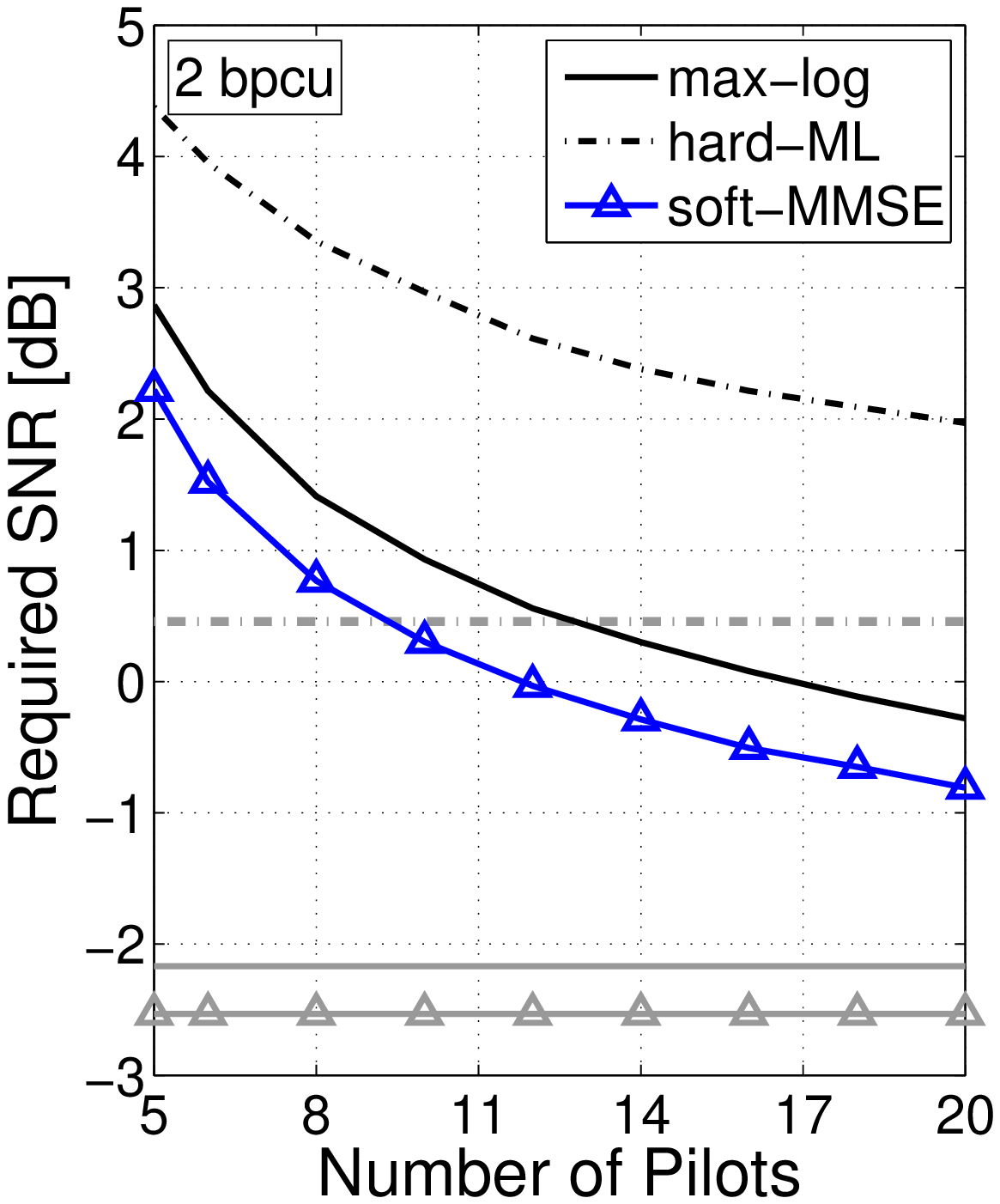}}
\subfigure[]{\includegraphics[scale=0.355]{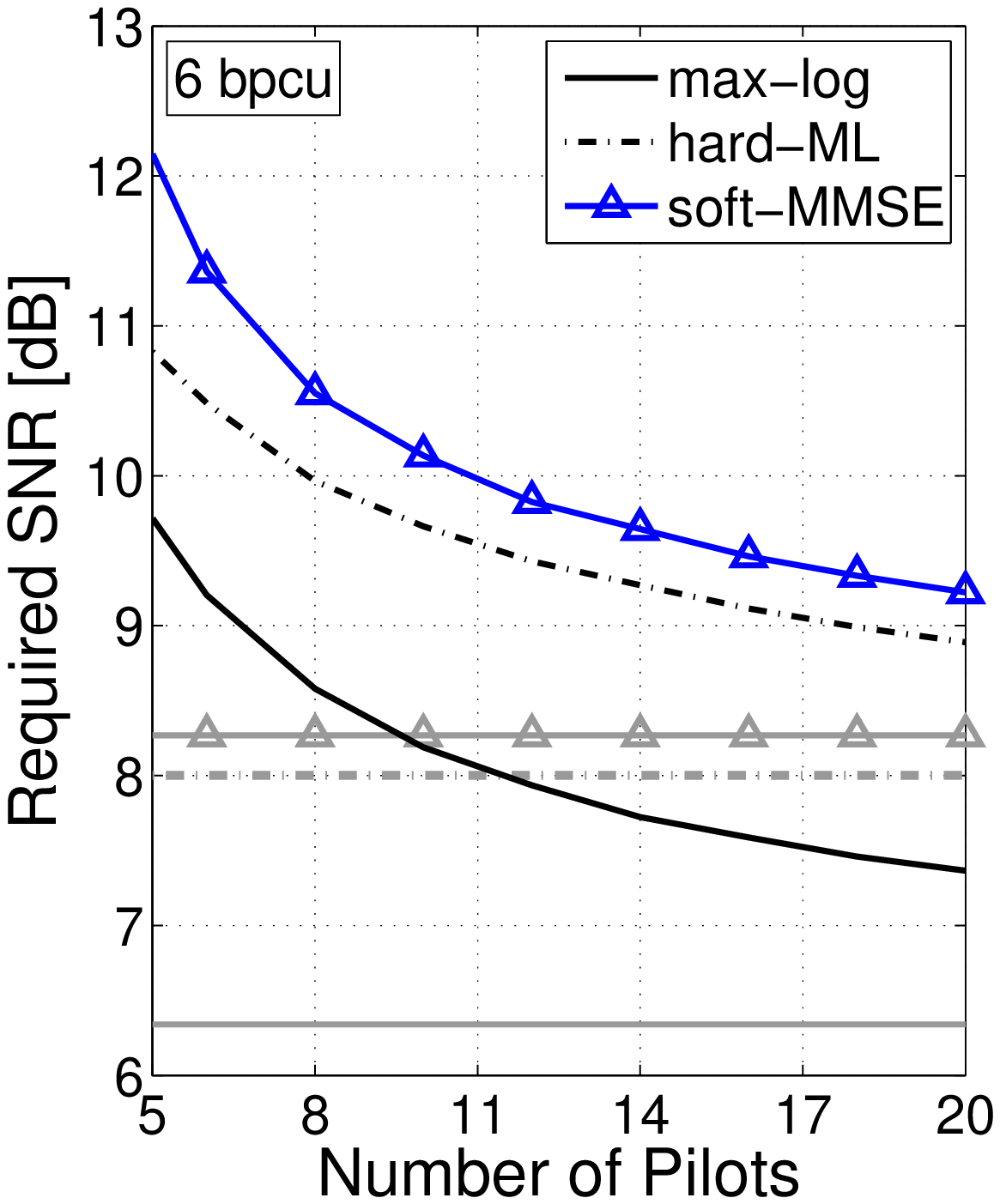}}
\vspace*{-3mm}
\caption{\label{fig:cap_iCSI_nvar}Impact of imperfect CSI on baseline demodulators: (a) capacity versus SNR for $\sNp\!=\!5$; (b), (c) required SNR versus training length $\sNp$ for a target rate of (b) $\sC=2$\,bpcu and (c) $\sC=6$\,bpcu ($4\!\times\!4$ MIMO, $4$-QAM, Gray labeling).}
\vspace*{-8mm}
\end{figure*}

Fig.~\ref{fig:cap_iCSI_nvar}(a) shows the maximum achievable rates versus SNR
for a fixed orthogonal training sequence of length $\sNp\!=\!5$ (the worst case with minimum amount of training).
It is seen that imperfect CSI causes a significant performance degradation of all three demodulators, e.g., at $4$\,bpcu the SNR losses are
$3.9$\,dB (max-log), 
$3.2$\,dB (hard ML), 
and $4$\,dB (soft MMSE). 
In this \NEW{worst case setup (minimum training length),} the performance advantage of soft MMSE over hard ML at low rates is slightly less pronounced; the intersection of hard ML and soft MMSE performance shifts from $5.8$\,bpcu (at an SNR of about $7.7$\,dB) for perfect CSI to $5$\,bpcu (at $9.4$\,dB) for imperfect CSI. However, the gap between soft MMSE and max-log is slightly larger at low rates, e.g., $0.7$\,dB at $2$\,bpcu.
The performance losses for all demodulators tend to be smaller at high rates, which may be partly attributed to the fact that the CSI becomes more accurate with increasing SNR. 
In general it can be observed that the performance loss of hard ML is the smallest while soft MMSE and max-log performance deteriorates stronger; 
unlike max-log and soft MMSE, 
hard ML does not use the noise variance and hence is more robust to estimation errors in $\ssigw$.


To investigate the impact of the amount of training, Fig.~\ref{fig:cap_iCSI_nvar}(b) and (c) depict the minimum SNRs required by the individual demodulators to achieve target rates of $2$\,bpcu and $6$\,bpcu, respectively, versus the training length $\sNp$. It is seen that for all demodulators, the required SNR decreases rapidly with increasing amount of training.
Yet, even for $\sNp=20$ there is a significant gap of $1$ to $2$\,dB to perfect CSI performance (indicated by horizontal gray lines with corresponding line style).
Here, soft MMSE consistently performs better than max-log and hard ML at $2$\,bpcu. 
In contrast, hard ML outperforms soft MMSE at $6$\,bpcu, especially for very small training durations.

\NEW{The results shown in Fig.~\ref{fig:cap_iCSI_nvar} correspond to a worst case scenario where both, channel and noise variance, are imperfectly known. 
Further capacity results, specifically for the case of imperfect channel knowledge but perfect noise variance 
and for other
demodulators discussed in Section \ref{sec:other}, are provided in \cite{Fertl:2009ac}}. These results generally show that imperfect receiver CSI degrades the performance 
throughout for all investigated demodulation schemes. An interesting observations is that---in the MIMO setup considered---the LSD with list size $\sLcand\ge 8$ consistently outperforms max-log 
for training duration $\sNp=5$ \NEW{\cite{Fertl:2009ac}}. 
For larger training durations, LSD with $\sLcand=8$ performs slightly poorer than but still very close to max-log.



\section{Quasi-static Fading}
\label{sec:slow}

In this section we provide a demodulator performance comparison for quasi-static fading MIMO channels based on
the outage probability $\sPout(\sRbar)$ in \eqref{eq:OUTAGE_PROBABILITY} and the $\epsilon$-capacity
$\sReps$ in \eqref{eq:EPSILON_CAPACITY}.
The setup considered ($4\times4$ MIMO with Gray-labeled $4$-QAM)
is the same as before 
apart from the spatially i.i.d.\ Rayleigh fading channel which now is assumed to be quasi-static.
The outage probability $\sPout(\sRbar)$ was measured
using $10^5$ blocks (affected by independent fading realizations),
each consisting of $10^5$ symbol vectors.
To keep the presentation concise,
we restrict to the baseline demodulators from Section \ref{sec:basic}.

\begin{figure*}
\centering
\subfigure[]{\includegraphics[scale=0.365]{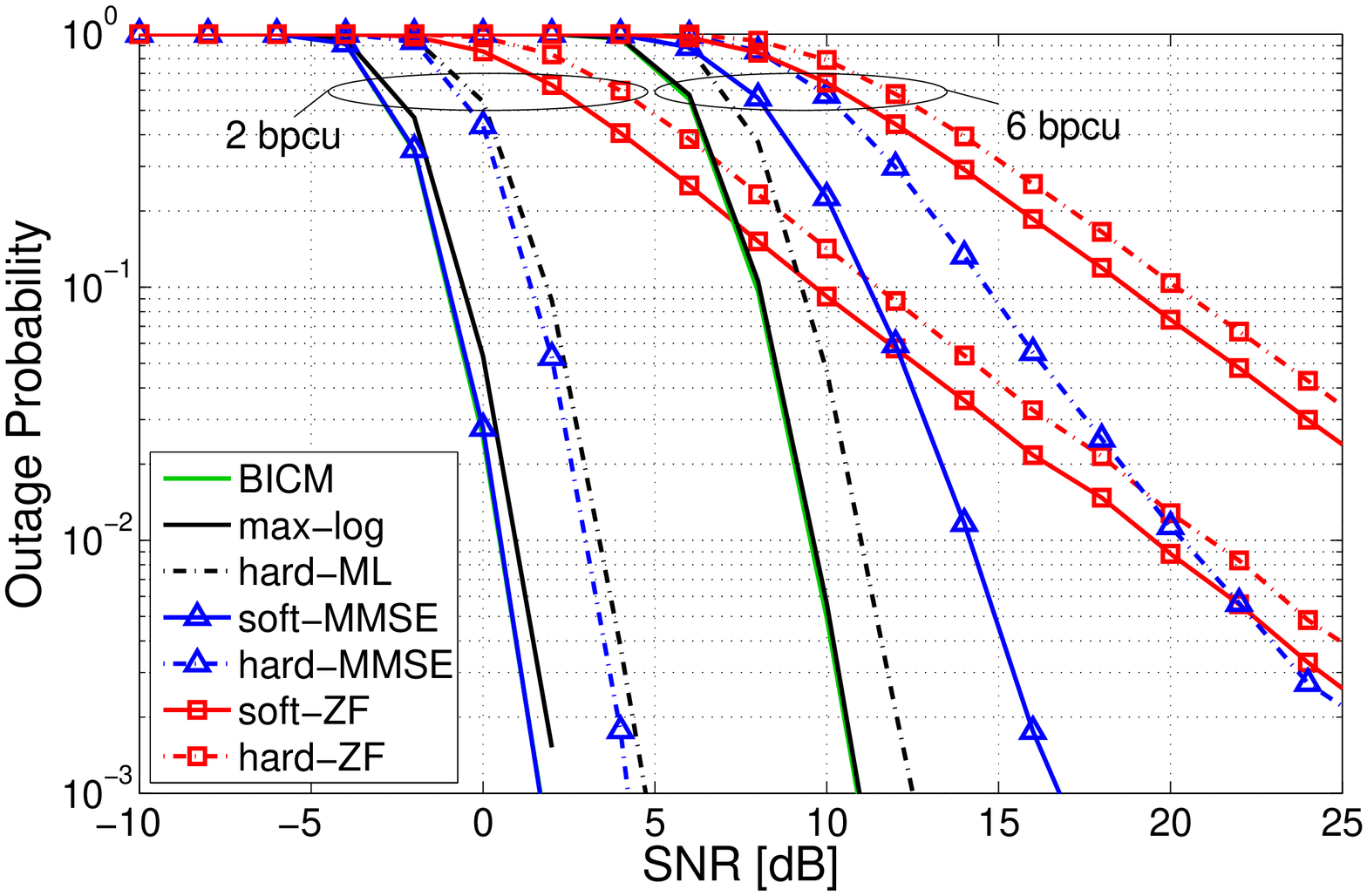}}
\hfill
\subfigure[]{\includegraphics[scale=0.365]{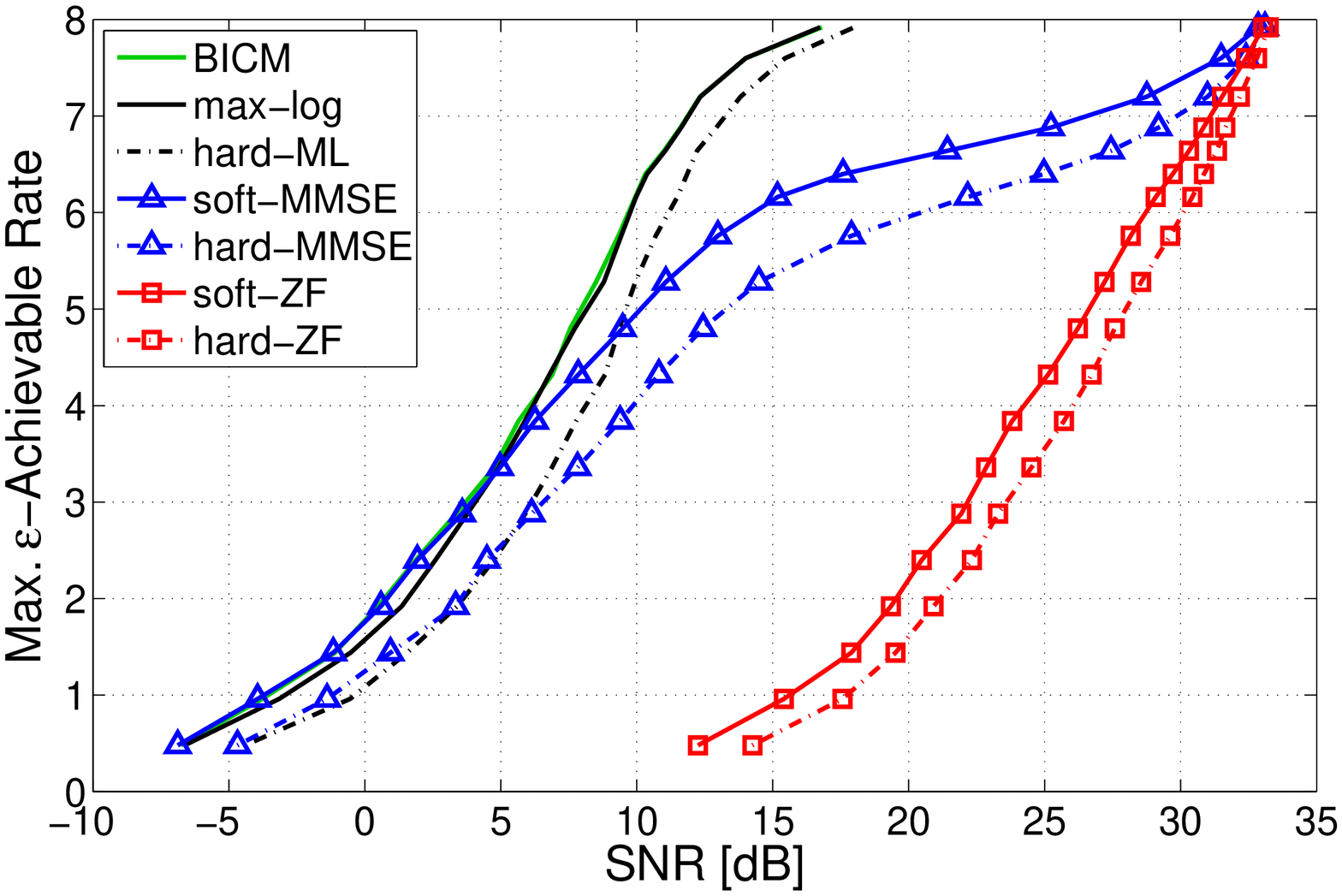}}
\vspace*{-4mm}
\caption{\label{fig:QS}Demodulator performance in quasi-static fading: (a) outage probability versus SNR for $\sRbar\!=\!2$\,bpcu and $\sRbar\!=\!6$\,bpcu, and (b) $\epsilon$-capacity versus SNR for $\epsilon\!=\!10^{-2}$ ($4\!\times\!4$ MIMO, $4$-QAM, Gray labeling).}
\vspace*{-7mm}
\end{figure*}

Fig.~\ref{fig:QS}(a) shows the outage probability versus SNR $\sSNR$
for target rates of $\sRbar\!=\!2$\,bpcu and $\sRbar\!=\!6$\,bpcu.
For $\sRbar\!=\!2$\,bpcu, 
soft MAP demodulation (labeled `BICM' for consistency with previous sections) and soft MMSE demodulation exactly coincide and outperform max-log demodulation by about $0.5$\,dB. In this low-rate regime, max-log performs about $2.5$\,dB better than hard ML. 
While max-log, hard ML, and soft MMSE demodulation all achieve full diversity (cf.\ the slope of the corresponding outage probability curves), soft and hard ZF only have diversity order one, resulting in a huge performance loss (almost $19$\,dB and $20.5$\,dB at $\sPout(\sRbar)=10^{-2}$, respectively).
At $\sRbar\!=\!6$\,bpcu the situation is quite different: here, max-log coincides with soft MAP and hard ML looses only $1.4$\,dB (again, those three demodulators achieve full diversity).
Hard and soft MMSE deteriorate at this rate and loose all diversity. 
At $\sPout(\sRbar)=10^{-2}$, the SNR loss of soft MMSE and soft ZF relative to max-log equals about $4.4$\,dB and $19$\,dB, respectively. 

The degradation of soft MMSE with increasing rate is also visible in Fig.~\ref{fig:QS}(b), which shows
$\epsilon$-capacity versus SNR for an outage probability of $\epsilon=10^{-2}$. 
The $\epsilon$-capacity qualitatively behaves similar as the ergodic capacity
(cf.~Fig.~\ref{fig:basic_gray}(a)): at low rates soft MMSE outperforms hard ML 
(by up to $2.8$\,dB for rates less than $4.7$\,bpcu)
while at high rates it is the opposite way. Furthermore, for low rates soft MMSE essentially coincides with soft MAP whereas at high rates it approaches soft ZF performance.

\NEW{
We note that a similar rate-dependent performance of MMSE demodulation has been observed in \cite{Hedayat:2007aa,Kumar:2009aa}. There it was shown that with coding across the antennas, 
the diversity order of MMSE equalization 
equals $\sMt\sMr$ at low rates 
and $\sMr-\sMt+1$ at high rates; in contrast, ZF equalization always achieves a diversity of only $\sMr-\sMt+1$. 
These results, which are interpreted in detail in \cite[Section IV]{Kumar:2009aa},
match well our observations that the MMSE demodulator looses diversity for rates larger than $5$\,bpcu 
(see Fig.~\ref{fig:QS} and \cite{Fertl:2009ac}).}


A comparison of Fig.~\ref{fig:QS}(b) and Fig.~\ref{fig:basic_gray}(a) suggests that there is a connection between the diversity of a demodulator in the quasi-static scenario and its SNR loss relative to optimum demodulation in the ergodic scenario.
For all rates (SNRs), the max-log and hard ML demodulator both achieve constant (full) diversity in the quasi-static regime and maintain a roughly constant gap to soft MAP in the ergodic scenario.
In contrast, with MMSE demodulation the diversity order 
in the quasi-static case and the SNR gap to soft MAP in the ergodic scenario both deteriorate with increasing rate/SNR.
\section{Key Observations and Design Guidelines}
\label{sec:practic}
Based on the previous results, we summarize 
key observations and provide 
system design guidelines.

\begin{figure}
\centering
\subfigure[]{\includegraphics[scale=0.365]{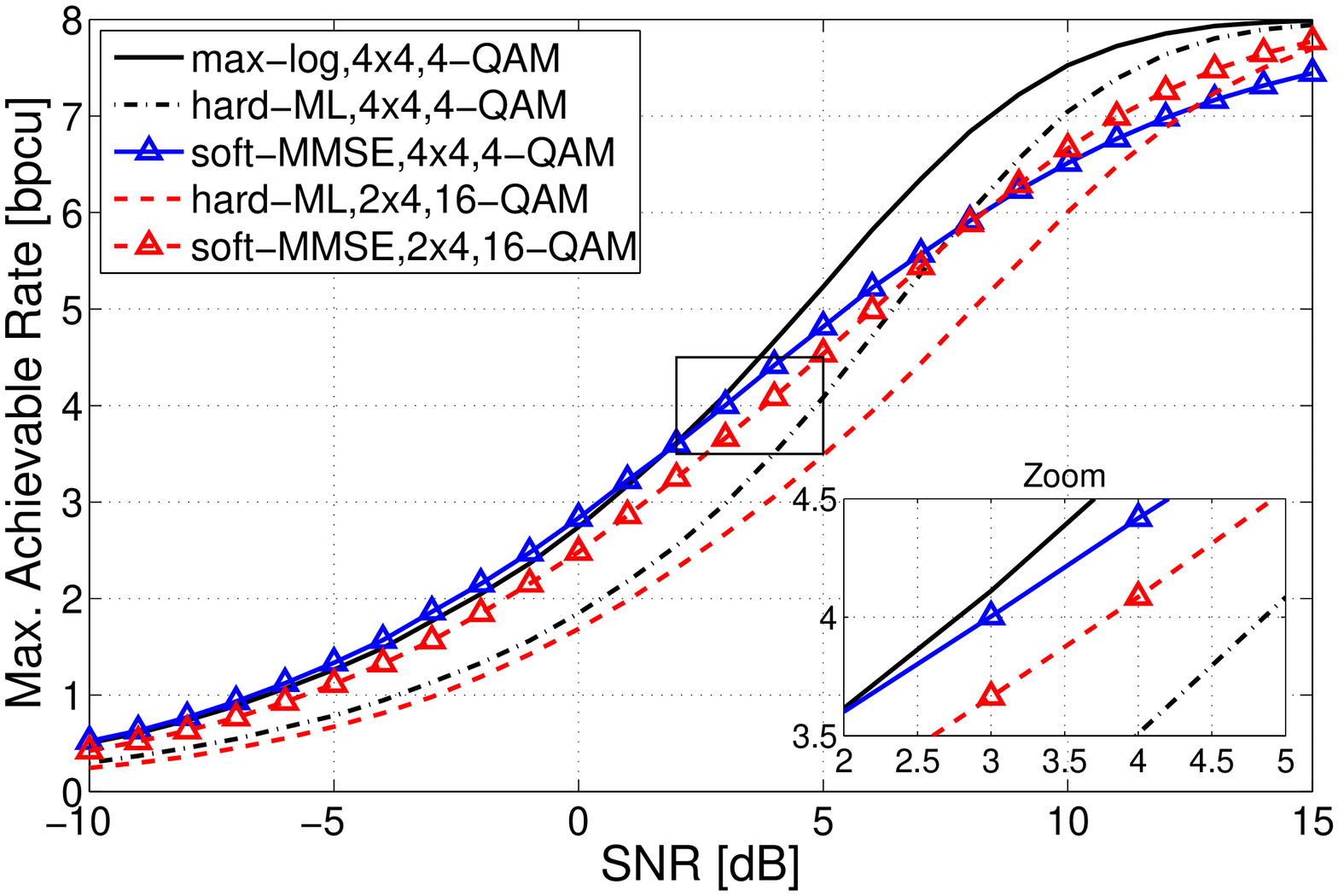}}
\hfill
\subfigure[]{\includegraphics[scale=0.365]{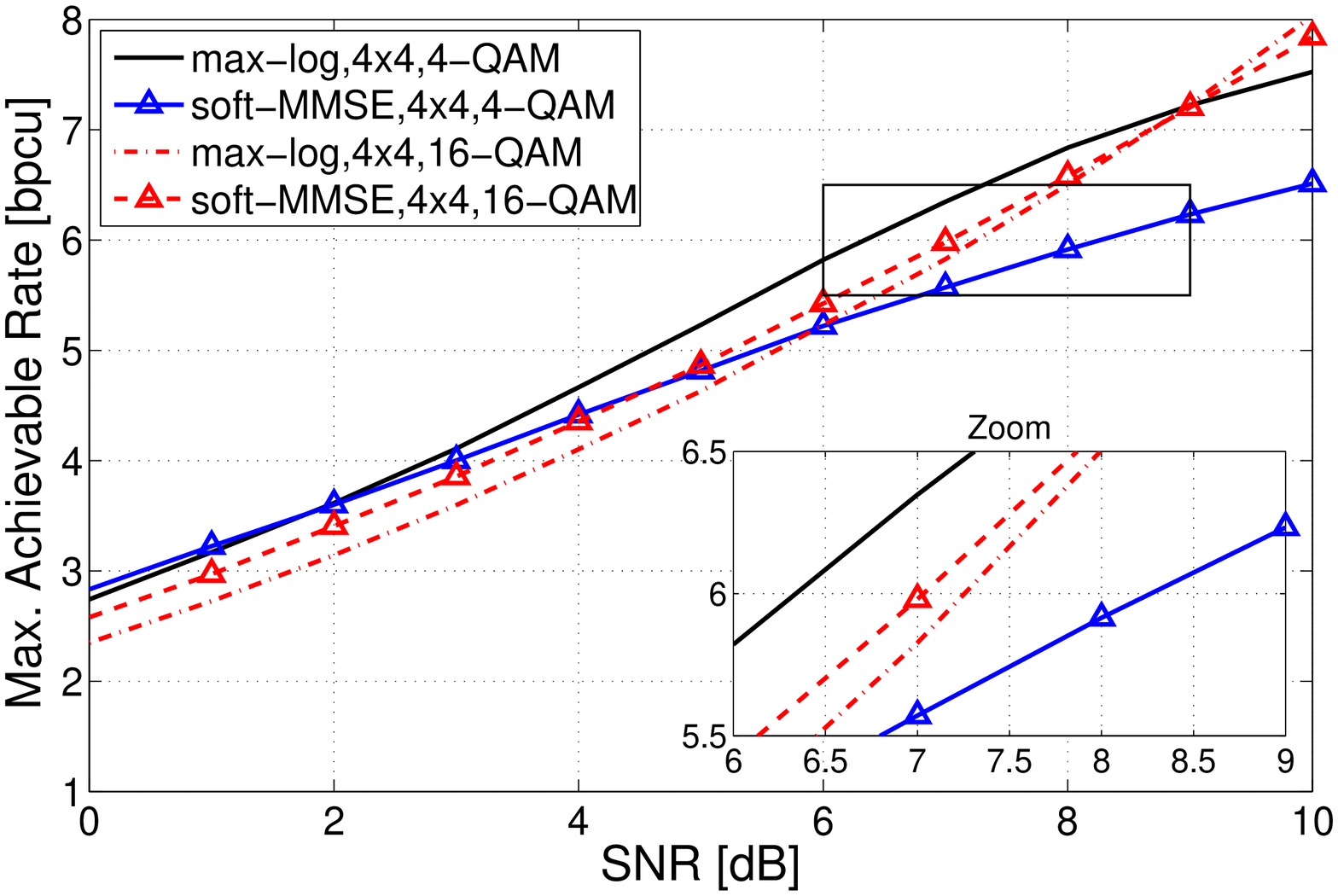}}
\vspace*{-4mm}
\caption{\label{fig:practic}System capacity of baseline demodulators for (a) a $4\!\times\!4$ MIMO system using $4$-QAM and a $2\!\times\!4$ MIMO system using $16$-QAM, and (b) a $4\!\times\!4$ MIMO system using $4$-QAM and $16$-QAM
(Gray labeling in all cases).}
\vspace*{-7mm}
\end{figure}

Soft MMSE demodulation approaches BICM capacity and outperforms max-log at low rates, both in the ergodic and the quasi-static regime and for various system configurations (see also \cite{Fertl:2009ac}).
Moreover, soft MMSE is very attractive since it has the lowest complexity
among all soft demodulators that we discussed.
Therefore, soft MMSE demodulation is arguably the demodulator of choice when designing MIMO-BICM systems with outer codes of low to medium rate. 
Since soft ZF 
performs consistently poorer than soft MMSE at the same computational cost, 
there appears to be no reason to prefer soft ZF in practical implementations.
The case for soft MMSE is particularly strong  
for asymmetric MIMO systems (i.e., $\sMt\!<\!\sMr$), where 
it performs close to BICM capacity for {\em all} rates.
Fig.~\ref{fig:practic}(a) compares a $4\times4$ MIMO system using $4$-QAM (system I) and a $2\times4$ system using $16$-QAM (system II), both using Gray labeling and achieving $\slambda\!=\!8$.
Whereas at low rates soft MMSE demodulation performs better with system I than with system II, it is the other way around for high rates. For example, at $7$\,bpcu system II requires $1.1$\,dB less SNR than system I, in spite of using fewer active transmit antennas. This observation is of interest when designing MIMO-BICM systems with adaptive modulation and coding. Specifically, with soft MMSE demodulation system I is preferable below $6$\,bpcu, whereas above $6$\,bpcu it is advantageous to deactivate two transmit antennas and switch to the $16$-QAM constellation (system II).
We note that with max-log and hard ML demodulation, system II performs consistently worse than system I.
The only regime where soft MMSE suffers from a noticeable performance loss is symmetric systems at high rates (both, in the ergodic and quasi-static scenario). 
In the high-rate regime, hard and soft SDR are the only low-complexity demodulation schemes that are able to achieve 
hard ML and max-log performance, respectively. 
These observations apply also in the case of imperfect CSI.
This suggests that hard and soft SDR demodulation are preferable over hard ML and max-log demodulation.
With perfect CSI, also LSD, bit-flipping demodulation, and soft $\slinf$-norm demodulation come reasonably close to max-log.
The LSD has the additional advantage of being able to trade off performance for complexity reduction.
Furthermore, note that for system I hard SDR demodulation (which approaches hard ML performance) 
outperforms most suboptimal soft demodulators for rates larger than $6$\,bpcu.

\NEW{The above discussion suggests that in order to achieve a given target information rate, it may be
preferable to adapt the number of antennas and the symbol constellation in a system with a low-rate code and a low-complexity MMSE demodulator instead of using a high-rate code and a computationally expensive non-linear demodulator. 
Such a design approach has recently been advocated also in \cite{Kumar:2009aa}. 
While RF complexity may be a limiting factor with respect to antenna number, increasing
the constellation size is inexpensive.
Fig.~\ref{fig:practic}(b) compares soft MMSE and max-log for a $4\times4$ MIMO system with Gray-labeled $4$-QAM ($\slambda\!=\!8$) and $16$-QAM ($\slambda\!=\!16$).
Below 3.5\,bpcu, soft MMSE demodulation with 4-QAM is optimal; at higher rates, 
switching to $16$-QAM allows the MMSE demodulator to perform within about $0.7$\,dB of
max-log with 4-QAM while increasing the soft-MMSE complexity only little. 
}

In case of imperfect receiver CSI, 
the performance of all 
demodulators deteriorates significantly (see also \cite{Fertl:2009ac}),
i.e., all capacity curves are shifted to higher SNRs (depending on the amount of training).
Demodulators that take the noise variance into account require somewhat more training.
\NEW{An exception is the LSD which can outperform max-log in case of poor channel and noise variance estimates \cite{Fertl:2009ac}.}

We conclude that at low rates linear soft demodulation is 
generally preferable due to its very low computational complexity. At high rates non-linear demodulators 
perform better, even when they deliver hard rather than soft outputs.
If complexity is not an issue, soft SDR demodulation is advantageous since it approaches max-log performance and is largely superior to all other demodulators 
over a wide range of system parameters and SNRs.
A notable exception is the low-complexity SoftIC demodulator which for some system configurations has the potential to outperform soft SDR (and max-log) at low rates.

\vspace*{-2mm}
\section{Conclusion}
\vspace*{-2mm}
\label{sec:conclusion}
We provided a comprehensive performance assessment and comparison of soft and hard demodulators for non-iterative MIMO-BICM systems. Our comparison is based on the information-theoretic notion of {\em system capacity\/},
which can be interpreted as the maximum achievable rate of the equivalent ``modulation'' channel that comprises modulator, physical channel, and demodulator. As a performance measure, system capacity has the main advantage of being independent of any outer code. Extensive simulation results show that a universal demodulator performance ranking is {\em not} possible and that the demodulator performance can depend strongly on the rate (or equivalently the SNR) at which a system operates. In addition to ergodic capacity results, we investigated the non-ergodic fading scenario in terms of outage probability and $\epsilon$-capacity and analyzed the robustness of certain demodulators under imperfect channel state information. Our observations provide new insights into the design of MIMO-BICM systems (i.e., choice of demodulator, number of antennas, and symbol constellation).
Moreover, our approach sheds light on issues 
that have not been apparent in the previously prevailing BER performance comparisons for specific outer codes. For example, a key observation is that with low-rate outer codes soft MMSE is preferable over other demodulators since it has low complexity but close-to-optimal performance.

\vspace*{-3mm}
\appendices

\NEW{
\section{Measuring Mutual Information}
\label{app:estimate-MI}
\vspace*{-1mm}
Evaluating the mutual information in \eqref{eq:measure-exact} involves the
conditional LLR distributions $\sprob(\shLLRj|\scj)$. We approximate these pdfs 
by histograms obtained via Monte Carlo simulations.
To achieve a small bias and variance of the mutual information estimate, the number 
and the size of the histogram bins as well as the sample size need to 
be carefully balanced \cite{Paninski:2003aa}.
Instead of LLRs, we use the bit probabilities 
\begin{equation*} 
\vspace*{-1mm}
\spdom_\sj=\frac{1}{1+\sexp^{-\shLLRj}}\,\in\,[0,1].
\end{equation*}
Since the LLRs $\shLLRj$ and $\spdom_\sj$ are in one-to-one correspondence (cf.~\eqref{eq:exact-LLR}), the mutual information of the equivalent modulation channel equals that of the channel characterized by the conditional pdf $\sprob(\spdom_\sj|\scj)$; the latter has the advantage of being easier to approximate by a histogram with uniform bins. 
By performing Monte Carlo simulations in which $\sN$ code bits, the noise, and the channel are
randomly generated, we obtain a histogram with $\sNbins$ bins which 
is characterized by the uniform bins $\big[\tfrac{k-1}{K},\tfrac{k}{K}\big]$, $k=1,\dots,K$, and the 
associated conditional relative frequencies $\shistkb$ (i.e., the normalized number of 
probabilities $\spdom_\sj$ lying in the $k$th bin conditioned on $\scj=\sbit$).
The mutual information in \eqref{eq:measure-exact} is then approximated as 
\begin{equation} \label{eq:measure-hist}
\sC \approx \sChat = \slambda \hspace*{.3mm} - \hspace*{.3mm} \sum_{\sj=1}^{\slambda} \sum_{\sbit=0}^1 \sum_{\sbini=1}^\sNbins \frac{1}{2}\shistkb \log_2 \!\frac{\sum_{\sbit'=0}^1 \shistkbt}{\shistkb}\,.
\end{equation}
%
The accuracy of this approximation depends 
i) on the number $\sNbins$ of histogram bins (this determines the discretization error)
and
ii) on the number $\sN$ of samples (code bits) used to estimate the histogram
(this determines the bias and variance of $\shistkb$ and hence of $\sChat$).
Specifically, the bias and the variance of $\sChat$ can be bounded as (see \cite{Paninski:2003aa})
\begin{equation} \label{eq:MI-bias}
0 \;\le\; 
 \sE\{\sChat\}-\sC_\text{Q}
\;\le\;
\log_2\!\bigg(1+\frac{\sNbins\!-\!1}{\sN}\bigg),\qquad
 \qquad 
\sE\Big\{\big(\sChat-\sE\{\sChat\}\big)^2\Big\}
\,\le \;
\frac{(\log_2 \sN)^2}{\sN}\,.
\end{equation}
Here, $\sC_\text{Q}$ is the mutual information of the discretized channel, i.e., equal to
\eqref{eq:measure-hist} but with $\shistkb$ replaced by 
$\sProb\big(\spdom_\sj\!\in\!\big[\tfrac{k-1}{K},\!\tfrac{k}{K}\big]\,\big|\,\scj\!=\!\sbit\big)$.
Hence, the bias in \eqref{eq:MI-bias} quantifies the systematic error resulting from the
empirical estimation of the histograms. 
We conclude that a large number $\sNbins$ of bins is advantageous in order to keep the discretization error small; in view of \eqref{eq:MI-bias}, this necessitates a significantly larger number $\sN$ of samples 
($\sN\gg\sNbins$)
in order to achieve a small estimation bias. Large $\sN$ simultaneously ensures a small estimation variance. The price paid for accurate capacity estimates is computational complexity.

To design $\sNbins$ and $\sN$, we first evaluated the BICM capacity in \eqref{eq:BICM} by direct numerical integration using the known pdfs in \eqref{eq:prob-dens}; then we estimated the same capacity via Monte Carlo simulations as described above using the optimum soft MAP demodulator and increasingly larger $\sNbins$ and $\sN$ until the result was close enough to the capacity obtained by direct evaluation. 
Specifically, with $\sNbins=256$ and $\sN=10^5$ the estimation error was on the order of $10^{-4}$
over a large SNR range. These numbers were then used to estimate the mutual information for all
other demodulators.

}


%
\vspace*{-3mm}
\section*{Acknowledgment}
\vspace*{-2mm}
The authors thank Gottfried Lechner for kindly providing his LDPC decoder implementation.

%
\renewcommand{\baselinestretch}{1.05}
\bibliographystyle{IEEEtran}
\bibliography{thesis}


%








\end{document}